\documentclass[conference]{IEEEtran}
\usepackage{times}

\usepackage[ruled]{algorithm2e} 
\usepackage{url}
\usepackage[utf8]{inputenc}
\usepackage{soul} 
\usepackage{inputenc}
\usepackage{CJKutf8}
\usepackage{xspace}
\usepackage{subfig}
\usepackage{bbm}
\usepackage{array,longtable}
\usepackage{balance}
\usepackage{graphicx}
\usepackage{times}
\usepackage{crossreference}
\usepackage{tikz}
\usepackage{booktabs}
\usepackage{array}
\usepackage{adjustbox}
\usepackage{multirow}
\usepackage{hyphenat}

\usepackage{listings}

\usepackage{footnote}
\makesavenoteenv{tabular}
\makesavenoteenv{table}
\usepackage{color}

\def\totalAppIdsCrawled{{1,554,253}\xspace}

\def\numberAppsVPNPermissionBeforeOct2011{{16}\xspace}

\def\percentageofAppsWithRatingGT4Dg50k{{68}}
\def\percentageofAppswith1tracker{{70}\xspace}

\def\totalReviewsChecked{{20,008}\xspace}

\usepackage{etoolbox}
\makeatletter
\preto\lstlisting{\def\@captype{table}}
\makeatother
\usepackage{pifont}

\renewcommand{\footnotesize}{\fontsize{9}{10}\selectfont}

\providecommand{\ie}{\emph{i.e.,\xspace} }
\providecommand{\eg}{\emph{e.g.,\xspace} }

\newcommand{\etal}{\mbox{\emph{et al.\xspace}}}

\usepackage{flushend}

\newcolumntype{L}[1]{>{\raggedright\let\newline\\\arraybackslash\hspace{0pt}}m{#1}}
\newcolumntype{C}[1]{>{\centering\let\newline\\\arraybackslash\hspace{0pt}}m{#1}}

\usepackage{amssymb}
\usepackage{pifont}
\newcommand{\cmark}{\ding{51}}%
\newcommand{\custompara}[1]{\vspace{2mm}\noindent{\bf{#1}}}
\begin{document}

\title{A First Look at Mobile Ad-Blocking Apps}


\author{\IEEEauthorblockN{Muhammad Ikram}
\IEEEauthorblockA{Data61, CSIRO and UNSW\\
Sydney, Australia\\
Email: Muhammad.Ikram@data61.csiro.au}
\and
\IEEEauthorblockN{Mohamed Ali Kaafar}
\IEEEauthorblockA{Data61, CSIRO\\
Sydney, Australia\\
Email: Dali.Kaafar@data61.csiro.au}}

\maketitle

\begin{abstract}
Online advertisers, third party trackers and analytics services are constantly tracking user activities as they access web services through their web browsers or mobile apps. While, web browser plugins disabling and blocking Ads (often associated tracking/analytics scripts), e.g. AdBlock Plus\cite{abpplugin} have been well studied and are relatively well understood, an emerging new category of apps in the tracking mobile eco-system, referred as the mobile Ad-Blocking apps, received very little to no attention. With the recent significant increase of the number of mobile Ad-Blockers and the exponential growth {of} mobile Ad-Blocking apps' popularity, this paper aims to fill in the gap and study this new category of players in the mobile ad/tracking eco-system. 

This paper presents the first study of Android Ad-Blocking apps (or Ad-Blockers), analysing 97 Ad-Blocking mobile apps extracted from a corpus of more than 1.5 million Android apps on Google Play. While the main (declared) purpose of the apps is to block advertisements and mobile tracking services, our data analysis revealed the paradoxical presence of third-party tracking libraries and permissions to access sensitive resources {on users'} mobile devices, as well as the existence of embedded malware code within some mobile Ad-Blockers. We also analysed user reviews and found that even though a fraction of users raised concerns about the privacy and the actual performance of the mobile Ad-Blocking apps, most of the apps still attract a relatively high rating. 

\end{abstract}

\section{Introduction}
\label{sec:intro}

Online advertising is ubiquitous in today's digital economy. Embedded third-party tracking libraries in websites and mobile applications (apps in short) is common and perform a variety of functions ranging from enhancing user experience, social sharing to the monetisation of services by enabling targeted and location-based advertisements. Such integration has evolved into a tangled eco-system illustrated by the top10K Alexa websites each integrating on average more than 30 different third-party tracking services for user profiling and advertisement purposes~\cite{sEnglehardt2016}. Likewise, the popularity of mobile apps resulted into a thriving mobile tracking and advertising ecosystem that is perhaps more privacy-invasive, due to the ubiquitous nature of the mobile apps usage.

Sensitive user data, including contacts, locations, SMS and web history is readily accessible on mobile devices for apps geared with the relevant permissions, and represent a valuable asset for third-party advertisers and trackers, which in turn poses serious privacy and security risks beyond the discomfort that mobile ads displayed on rather small screens of smartphones might generate. Naturally, the mobile apps eco-system has recently witnessed the emergence of a new class of Ad-blocking\footnote{We use the term Ad-Blocking to refer to apps blocking both ads and tracking/analytics services.} tools, packaged as mobile apps, 
in popular mobile app stores such as Google Play. 

This paper presents the first characterisation study of Android mobile Ad-Blocking apps with a focus on security and privacy offered by these apps. In particular, we analyse the Android permissions mobile Ad-Blockers request and we perform static analysis of the code to investigate the presence of malware and third party tracking libraries. 

We collect and extract from a corpus of more than 1.5 million Android apps, 
97 mobile apps for which the name or the description suggest they enable to either block ads or to block trackers. We then manually check that the apps actually fall into the category of Ad-Blocking apps (cf. Section~\ref{sec:adbchracterization}). 

We use a set of tools to decompile the Ad-Blocking apps and analyse the source code of each of the mobile Ad-Blockers. We then inspect the apps to reveal the presence of third-party tracking libraries and sensitive permissions for critical resources on users' mobile devices. According to VirusTotal\footnote{An Anti-virus (AV) tools conglomerate, https://virustotal.com}'s classification, we observe instances of Ad-Blockers using excessive advertising, displaying full-screen advertisements windows and embedding malware in the source code (cf. Section~\ref{sec:sanalysis}).

This paper makes the following major contributions: 
\begin{itemize}
\item \textbf{Mobile Ad-Blockers identification and taxonomy:} We investigate a dataset of \totalAppIdsCrawled Android apps on Google Play and identify 97 Ad-Blocking apps. 
Based on the intended functionality and an inspection of the code, we provide a generic taxonomy of Ad-Blockers depending on the mechanisms used to block ads and the (corresponding) tracking/analytics services.
\item \textbf{Third-party user tracking:} We perform static analysis on mobile Ad-Blockers' source code. We observe that, albeit Ad-Blocking apps' claims to block ads and prevent tracking, 68\% of them still embed third-party tracking and ads libraries in their code, potentially leaking personal information to third-parties. We also observe that 24\% display ads. 
\item \textbf{Sensitive Permissions Access:} Our analysis reveal that 89\% of the Ad-Blockers request sensitive permissions to access critical resources such as user contacts, accounts, text messages, and user browsing history.
\item \textbf{Malware presence and inefficiency of Ad-Blockers:} According to VirusTotal scan reports, 13\% of the Ad-Blockers have malware presence in their source code with instances of spyware and adware to spy on {users'} behaviour. We analyse user reviews on Google Play to sense whether users expressed concerns about the security, the privacy, and the inefficiency--in term of not blocking ads and trackers--of the Ad-Blockers. Our analysis reveals that albeit users have publicly raised concerns in their app reviews yet some of the Ad-Blockers have high ratings and a significant number of installs. 
\end{itemize}

\section{Characterizing Ad-Blockers on Google Play}
\label{sec:adbchracterization}

We first briefly introduce mechanisms used by Mobile Ad-Blocking apps to block trackers and ads. Then, we describe our method for identifying Ad-Blockers on Google Play and show our characterisation of the collected Ad-Blockers.



\subsubsection{Android Ad-Blocking Mechanisms}
\label{subsec:adblockingmechanisms}
\begin{figure}[!ht]
\centering
\includegraphics[width=0.99\columnwidth, height=0.48\textwidth, keepaspectratio]{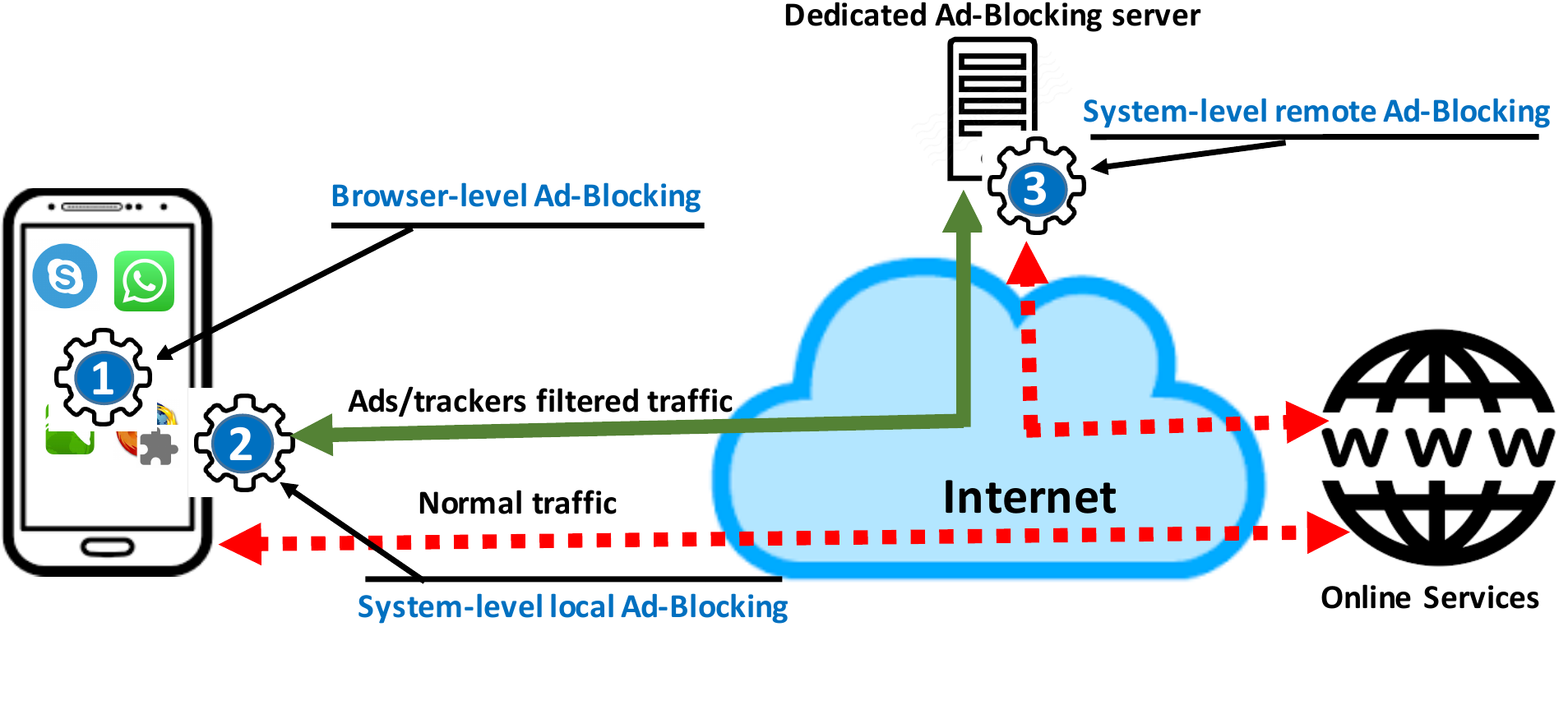}
\caption{Overview of Android's Ad-Blocking mechanisms: (1) \emph{brower-level Ad-Blocking}--browser with built-in functions or add-ons to block third-party advertisements and trackers on browsed webpages, (2) \emph{system-level local Ad-Blocking}--specialized Android apps which intercept all installed apps' traffic and filter-out advertisements and trackers, (3) \emph{system-level remote Ad-Blocking}--specialized Android apps which forward user traffic to dedicated servers for Ad-Blocking.}
\vspace{-0.2cm}
\label{fig:thirdparty_ads}
\end{figure}

Third-party advertisers and trackers use HTTP(s) to deliver Ads to end users. Ad-Blockers intercept and filter ads- and tracking-related traffic. To this end, Ad-Blockers can employ filters either as browsers' extensions (add-ons) or as a built-in functionality (\emph{Case 1} in Figure~\ref{fig:thirdparty_ads}) to block ads and trackers from the traffic generated while the user (or associated app) accesses the Internet. Ad-blockers can also implement mechanisms such as virtual private networks (VPN) tunnels\footnote{https://developer.android.com/reference/android/net/ VpnService.html} to intercept and filter ads-related traffic, either locally on mobile devices (Case \emph{2} in Figure~\ref{fig:thirdparty_ads}) or on remote servers (Case \emph{3} in Figure~\ref{fig:thirdparty_ads}), from all installed apps.

\subsubsection{Dataset in Use}


In order to identify as many Ad-Blockers as possible on Google Play, we implemented a Google Play crawler that takes advantage of two complementary seeds. First, we start with the top 100 apps for four Google Play categories likely to contain Ad-Blockers: Tools, Communication, Personality, and Productivity. Second, we use Google Play's search feature to find apps containing Ad-Blocking-related keywords. The keywords used are: ``Ad-Block'' (and variants including ``ads block'', ``ad block'', ``ads-blocking'', ``blocking-ads'' etc.), and ``privacy'', ``tracking-free'', ``Anti-tracking'' in their app descriptions. 
Our crawler follows a breadth-first-search approach for any other app considered as {\em ``similar''} by Google Play and for other apps published by the same developer. 
In total, we surveyed \totalAppIdsCrawled apps for a 4-week period in December 2016.

Android apps are typically written in Java code (and possibly with some additional native code). The overall Java code--implementing {the} intended functionality and third-party library for extended service or feature--is compiled to a {\tt .dex} file, containing compressed bytecode that runs in the Dalvik virtual machine. Android apps are distributed on marketplaces such as Google Play store as {\tt .apk} files, which bundle the {\tt .dex} code with the {app's} specification file {named {\tt AndroidManifest.xml}}. To download the apps' {\tt .apk} files ({\em aka} APKs) and other apps' metadata from Google Play (\eg app description, number of installs, developer information, user reviews and app rating), we use Google Play Unofficial Python API~\cite{unofficialg} for free apps and {the} Raccoon APK Downloader for paid apps~\cite{raccoon}. Finally, we use {\tt ApkTool} and {\tt dex2jar} to decompile and extract each app's source code and the corresponding {\tt AndroidManifest.xml}. 

We identified a total of 97 Ad-Blockers (87 free and 10 paid  apps) that match one or more relevant keywords in their app description. We manually checked their description to ensure they do belong to what users would consider as an Ad-Blocking tool or alternatively a tool to block tracking. For a baseline comparison (cf. Section~\ref{sec:sanalysis}), we also collected 500 randomly selected free non-Ad-Blocking apps--collected from Tools, Communication, Personality, and Productivity categories--from Google Play. 
In order to further analyse adblockers and reproduce our findings, the dataset and crawling scripts are available upon request.

%

\subsubsection{The Rise of Android Ad-Blockers}

We start our analysis by observing the increase of Ad-Blockers available for download on Google Play over time. Given that Google Play does not report the actual release date of the apps but their last update, we use the date of their first review as a proxy of their release date. For those Ad-Blockers without user reviews, we approximate the release date with the date of their last update. 

\begin{figure}[!ht]
\centering
\subfloat[]{\label{fig:apps_on_gplay} 
\includegraphics[width=0.99\columnwidth]{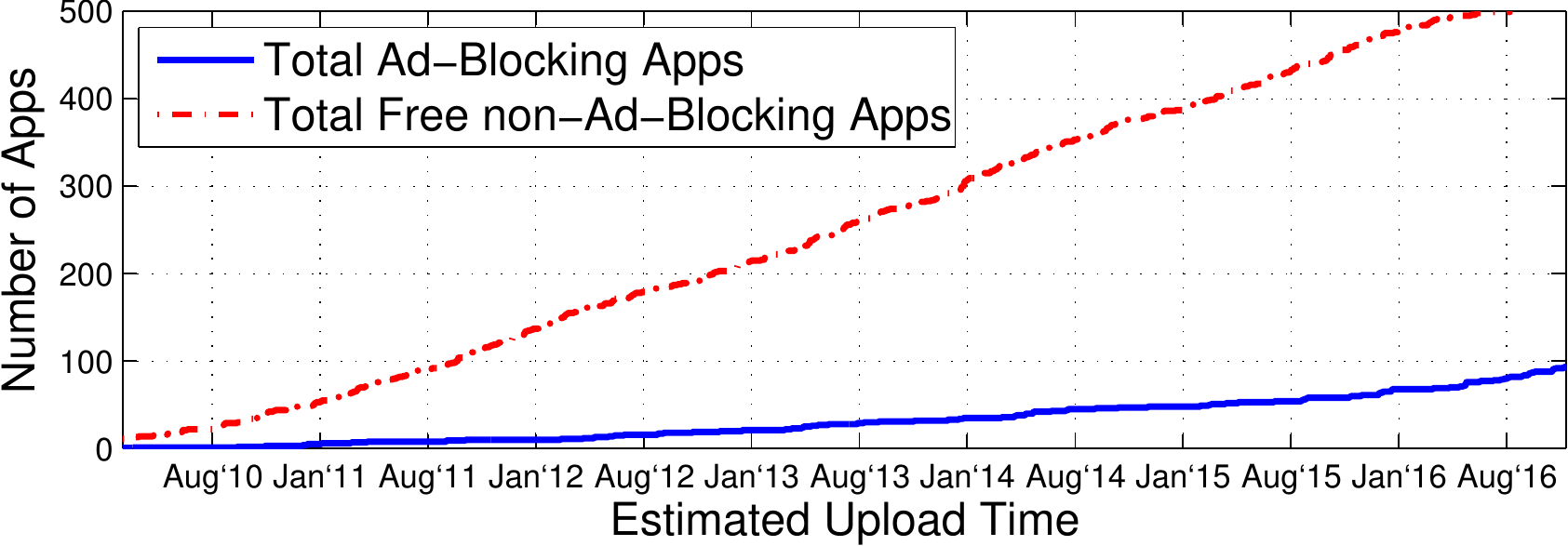}}

\subfloat[] {\label{fig:perceptionfig}
\includegraphics[width=0.99\columnwidth]{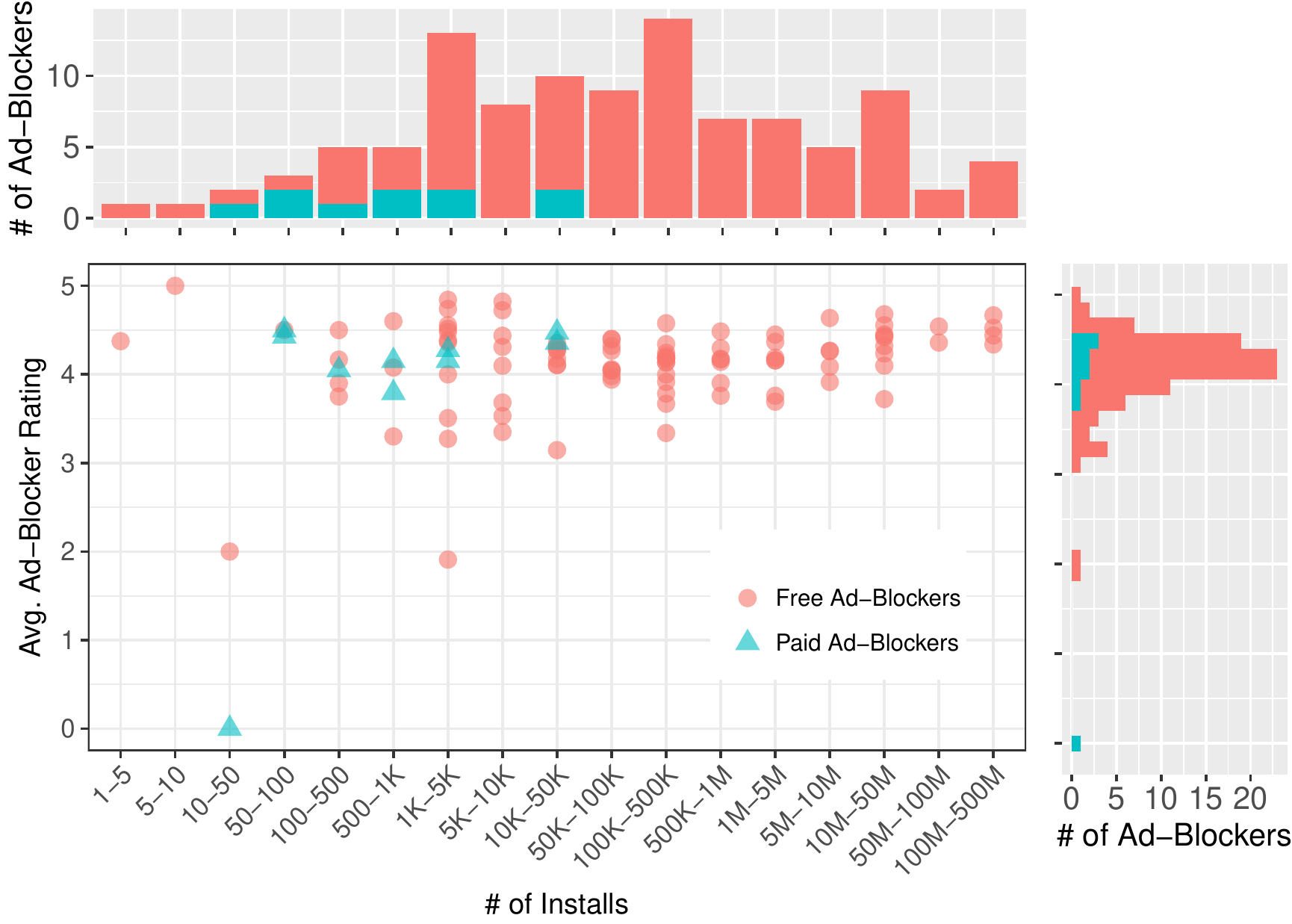}}
\vspace{-0.2cm}
\caption{Evolution of Ad-Blockers on Google Play: \textcolor{red}{(a)} Number of Ad-Blockers available on Google Play over time. \textcolor{red}{(b)} Distribution of app rating vs. installs per Ad-Blocker. }
\vspace{-0.4cm}
\label{fig:adblockers_evolution}

\end{figure}

Figure~\ref{fig:apps_on_gplay} shows the steady increase of Ad-Blocking apps listed on Google Play. During the 3-year period that spans between August 2012 and August 2015, the number of Ad-Blocking apps increased 3-fold. 
Overall, the analysed Ad-Blockers receive high user ratings: 41\% of the Ad-Blockers have more than 100K 
installs and \percentageofAppsWithRatingGT4Dg50k\% of them have 
at least a 4-star rating as shown in Figure~\ref{fig:perceptionfig}. 
We cannot tell whether the installs and the reviews are legitimate or if those ratings were actually acquired by the app developers, using apps' promotion services on underground marketplaces to promote the apps\footnote{https://www.seoclerk.com/Link-Building/192193/Test-Your-iOS-or-Android-Apps-On-Smartphone-And-Provide-Review-And-Rating}.

%

%


\custompara{Ad-Blockers' Classification.}

Given that Ad-Blockers block ads and trackers in a different fashion, we aim at categorising Ad-Blockers according to their intended functionality (or complementary features) and their Ad-Blocking mechanism. 
Unfortunately, Google Play's app categories (e.g., Tools and Communication) are too broad to capture the actual purpose and functionality of the app. Moreover, apps {may not include any detail on} their Ad-Blocking mechanisms in their descriptions on Google Play. 
To identify Ad-Blocking mechanisms of the analysed Ad-Blockers, we manually tested and categorised them into two classes, listed in Table~\ref{tab:classification}. 

\begin{table}[h]
\scriptsize
\centering
\begin{tabular}{L{4.5cm} C{3.3cm}}
  	\toprule
   	 {\bf App Category} &{\bf \% of Apps (N = 97) }  \\ 
    	\midrule
    Browser-level Ad-Blockers 	& 86\% \\
    System-level or VPN-based Ad-Blockers	& 14\% \\
    \bottomrule
	 \end{tabular}
	 \vspace{-0.2cm}
	\caption{Classification of Ad-Blockers (see Table~\ref{tab:alladblockers} for details).}
	\label{tab:classification}
	\vspace{-0.2cm}
\end{table}

\begin{figure}[!t]
\centering
\includegraphics[width=0.99\columnwidth, height=0.48\textwidth, keepaspectratio]{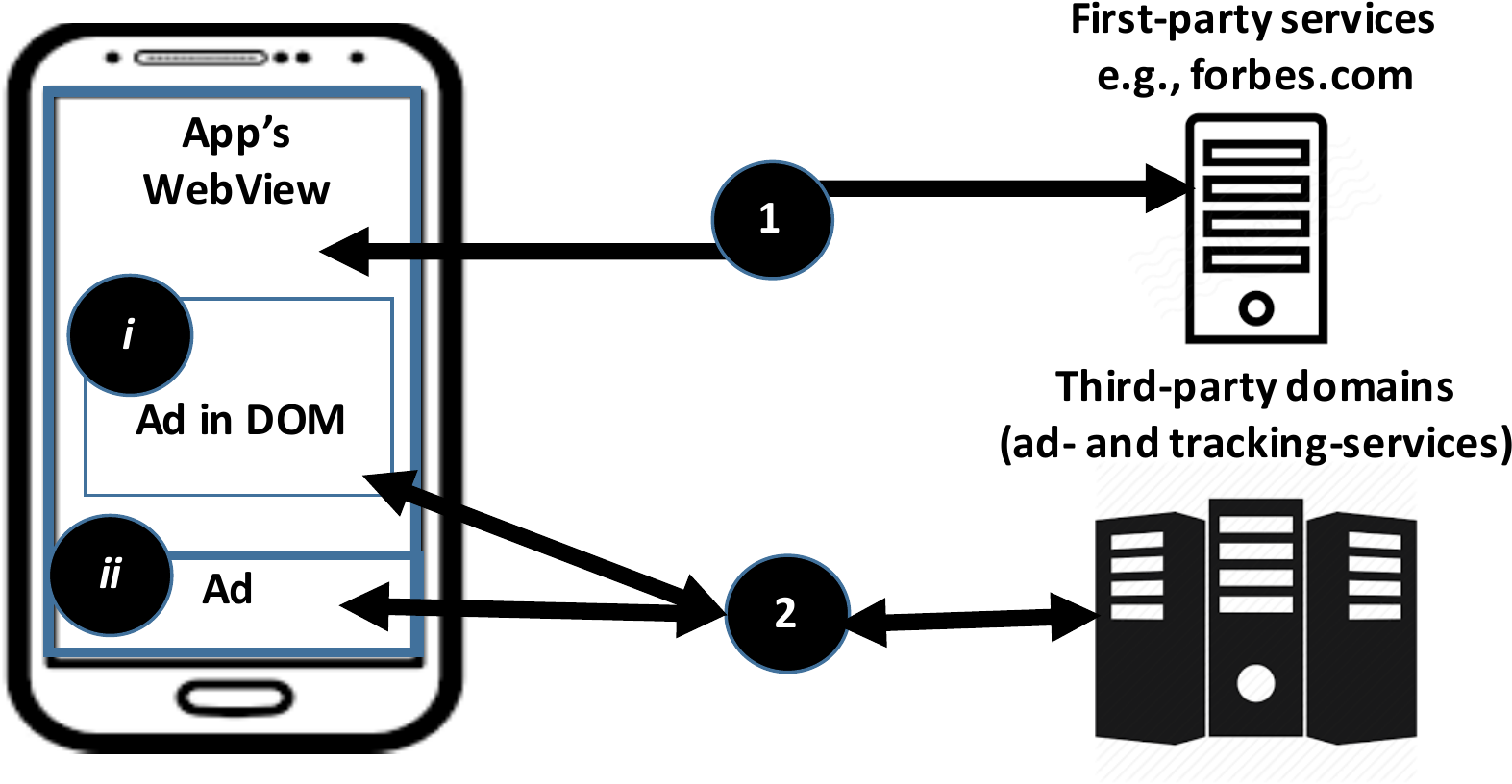}
\caption{Overview of ads displayed in Android apps. Here, an app accesses, \emph{step 1}, first-party webpage. Depending upon third-party libraries embedded in app's source code and third-party JavaScript programs included in the webpage's HTML source code, the app sends additional requests, \emph{step 2}, to third-party ad services. The app displays ads from third-party ad services either in In-App Ads-banner, \emph{Case ii}, in a browsed webpage, \emph{Case i}, or both. }
\label{fig:androidappmodel}
\end{figure}

We found that 86\% of the analysed Ad-Blockers(cf. Table~\ref{tab:alladblockers})
have built-in Ad-Blocking mechanism to block ads and trackers on a given webpage. These Ad-Blockers do not block In-App ads or Ad-banners (cf. Case (\emph{ii}) in Figure~\ref{fig:androidappmodel}).  On the other hand, 14\% of the apps create local or remote VPN-tunnel for blocking ads-related traffic, whether ads appeared in ad-banners or in a browsed webpage (cf. Case \emph{(i) \& (ii)} in Figure~\ref{fig:androidappmodel}), from all installed apps. 

\custompara{Geographical Distribution of Ad-Blockers.}

Here, we are interested in characterising the distribution and the popularity of Ad-Blockers per country and per geographical regions. 
To this end, we rely on the geographical popularity or \emph{rank} data from AppAnnie~\cite{appannie}. AppAnnie offers downloads or installs analytics to apps' developers and gathers apps' longitudinal usage from app markets such Google Play and iTunes. It uses developers' accounts to tracks apps' downloads from app markets and assigns numeric values or \textit{ranks} to represent apps' popularity in each geographical region or country. 
For each Ad-Blocker, we use AppAnnie API to  
obtain rank values per country.  
From the collected rank data, we map Ad-Blockers to countries and count the number of distinct Ad-Blockers per country. 
Moreover, we measure the median ranks to determine the popularity of Ad-Blockers per country. 

\begin{figure}[ht]
\centering
\includegraphics[width=0.99\columnwidth]{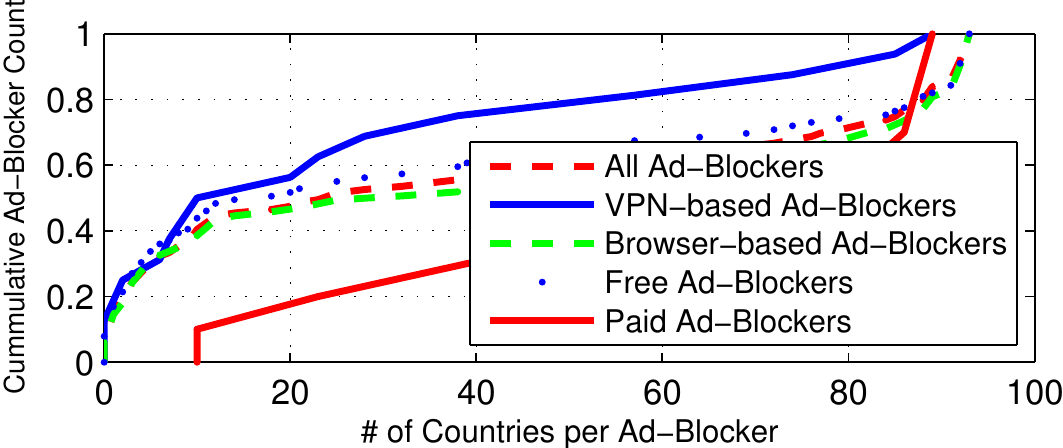}
\caption{Geographical usage (or installs) of Ad-Blockers.}
\label{fig:adblockers_gelocation}
\end{figure}

Figure~\ref{fig:adblockers_gelocation} shows the cumulative distribution of the number of countries with the number of distinct Ad-Blockers according to AppAnnie countries rank dataset. We observe a significant difference in the geographical coverage among the classes of Ad-Blockers. The distribution suggests that paid Ad-Blockers have more geographically scattered downloads around the globe: 80\% of paid Ad-Blockers have downloads in more than 20 countries where as the 56\% of the free counterparts have their users located in less than 20 different countries. The figure also reveals that browser-level Ad-Blockers have a higher geographical coverage in terms of installs when compared to systems-level or VPN-based Ad-Blockers. 

We also observe a significant difference in the number of Ad-Blockers per geographical region. 
Several countries including China, Macedonia, and Lativa have 
diverse set of free Ad-Blockers--each has 56\% of distinct Ad-Blockers. 
Maxthon Browser, a browser-based free Ad-Blocker, has more than 10M installs in 93 different countries, suggesting its global popularity. On the other hand Piggy Browser, {also a free} Ad-Blocker, has 1K installs only in a single country, Japan.


\begin{figure}[ht]
\centering
\subfloat[All Ad-Blockers]{\label{fig:over_rank} 
\includegraphics[width=0.99\columnwidth]{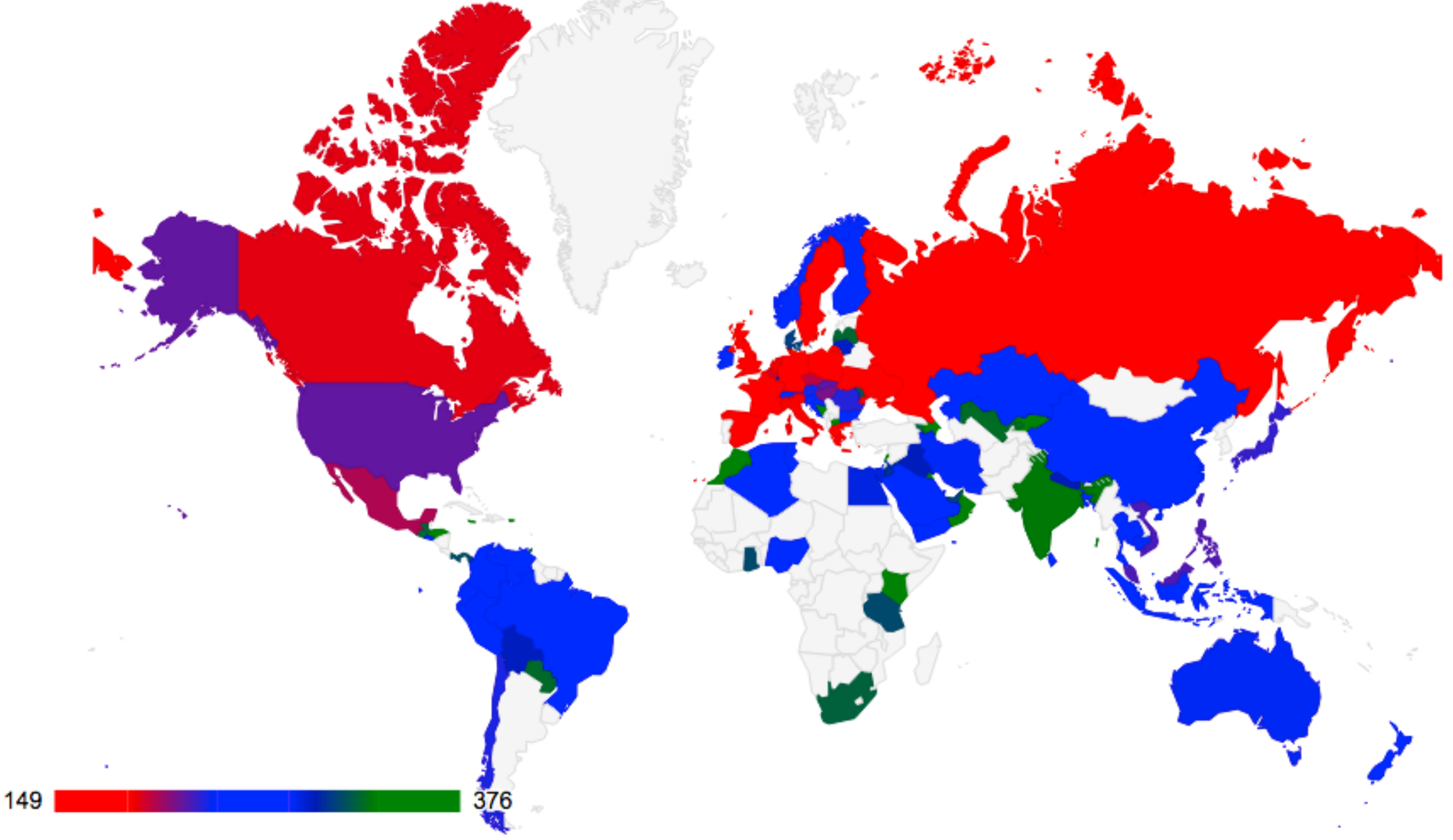}}\hspace{2em}
\subfloat[Paid Ad-Blockers] {\label{fig:paid_adb}
\includegraphics[width=0.49\columnwidth]{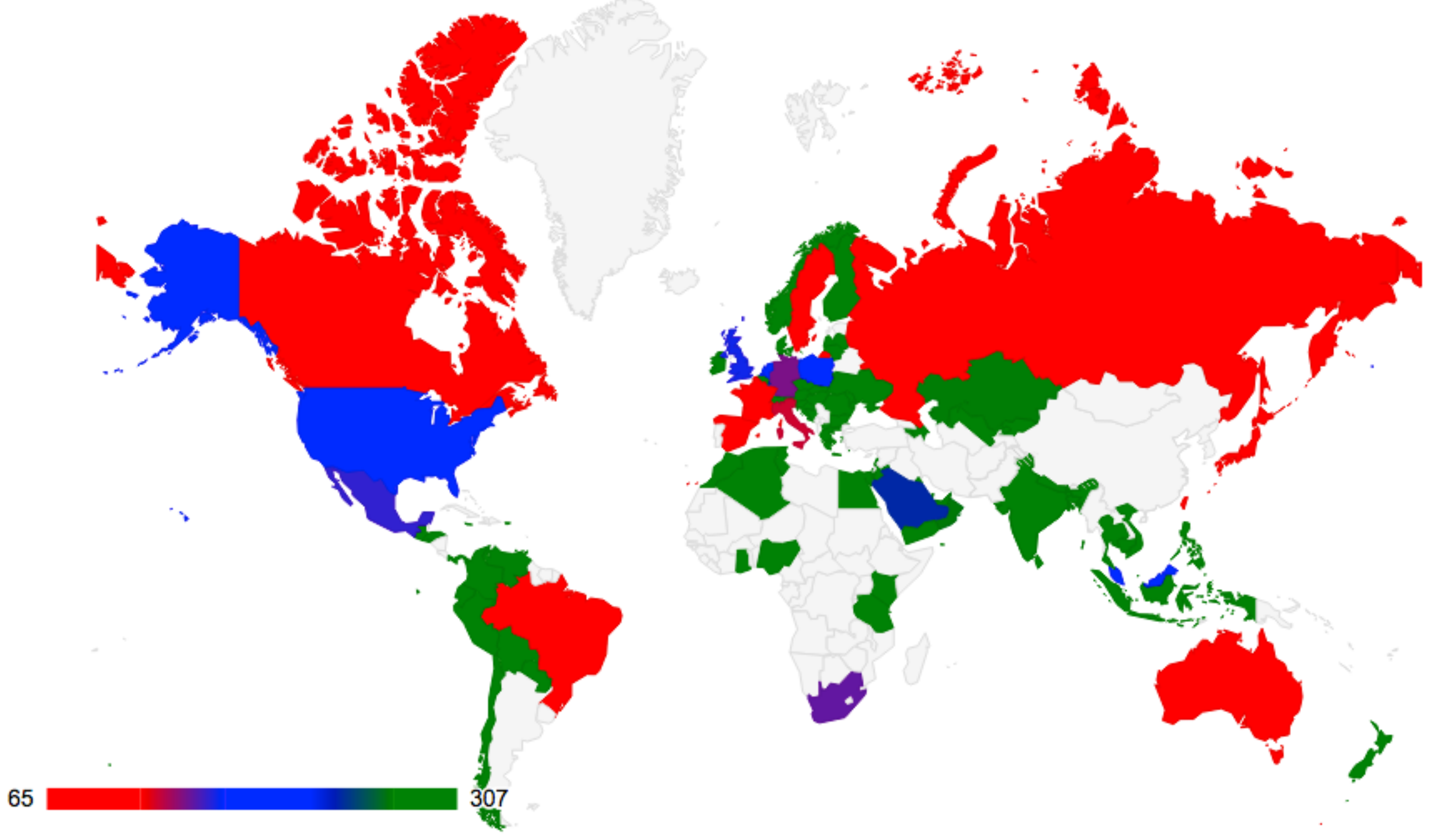}}
\subfloat[Free Ad-Blockers] {\label{fig:free_adb}
\includegraphics[width=0.49\columnwidth]{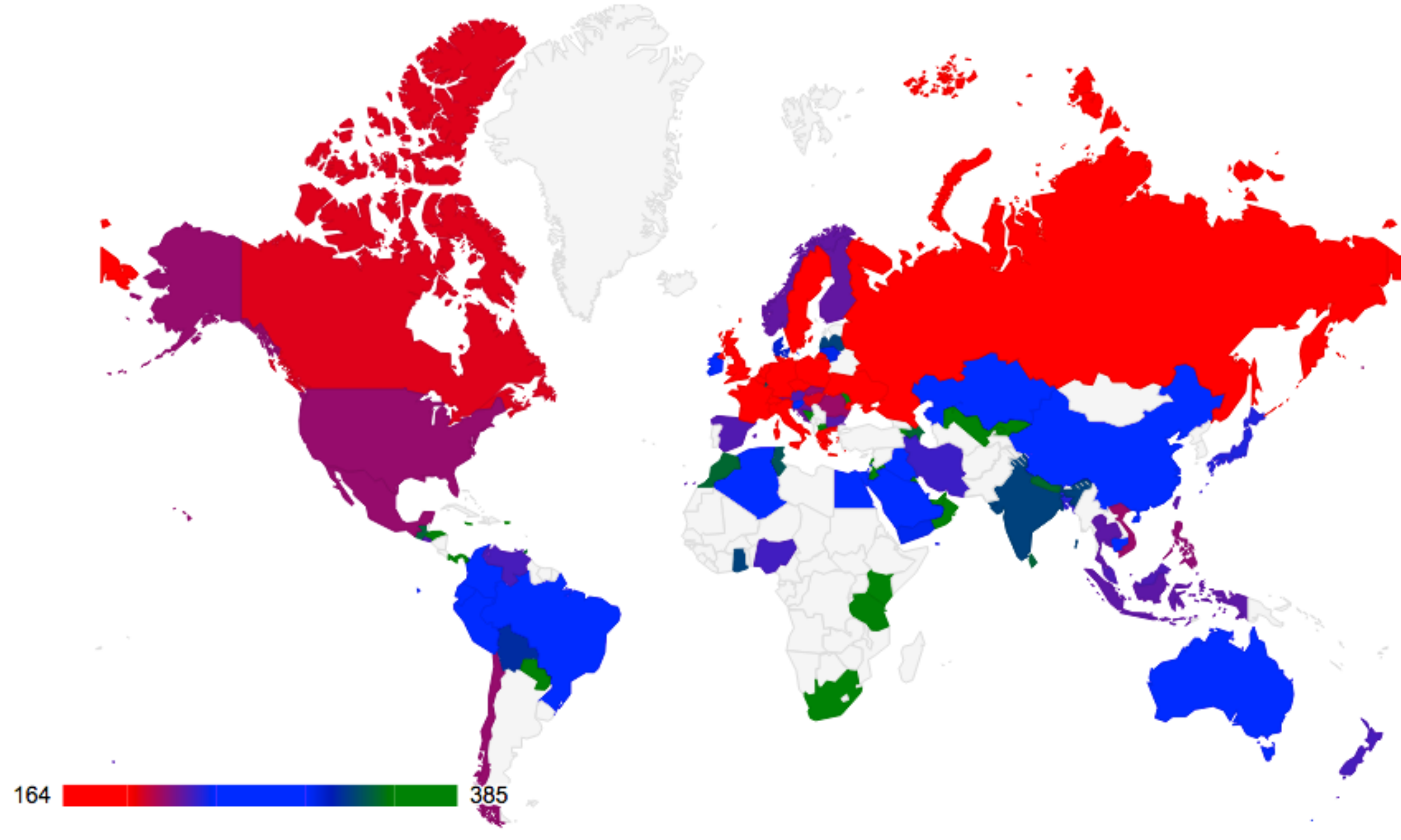}}\hspace{2em}
\subfloat[System-level(VPN-based) Ad-Blockers] {\label{fig:vpn_based_adb}
\includegraphics[width=0.49\columnwidth]{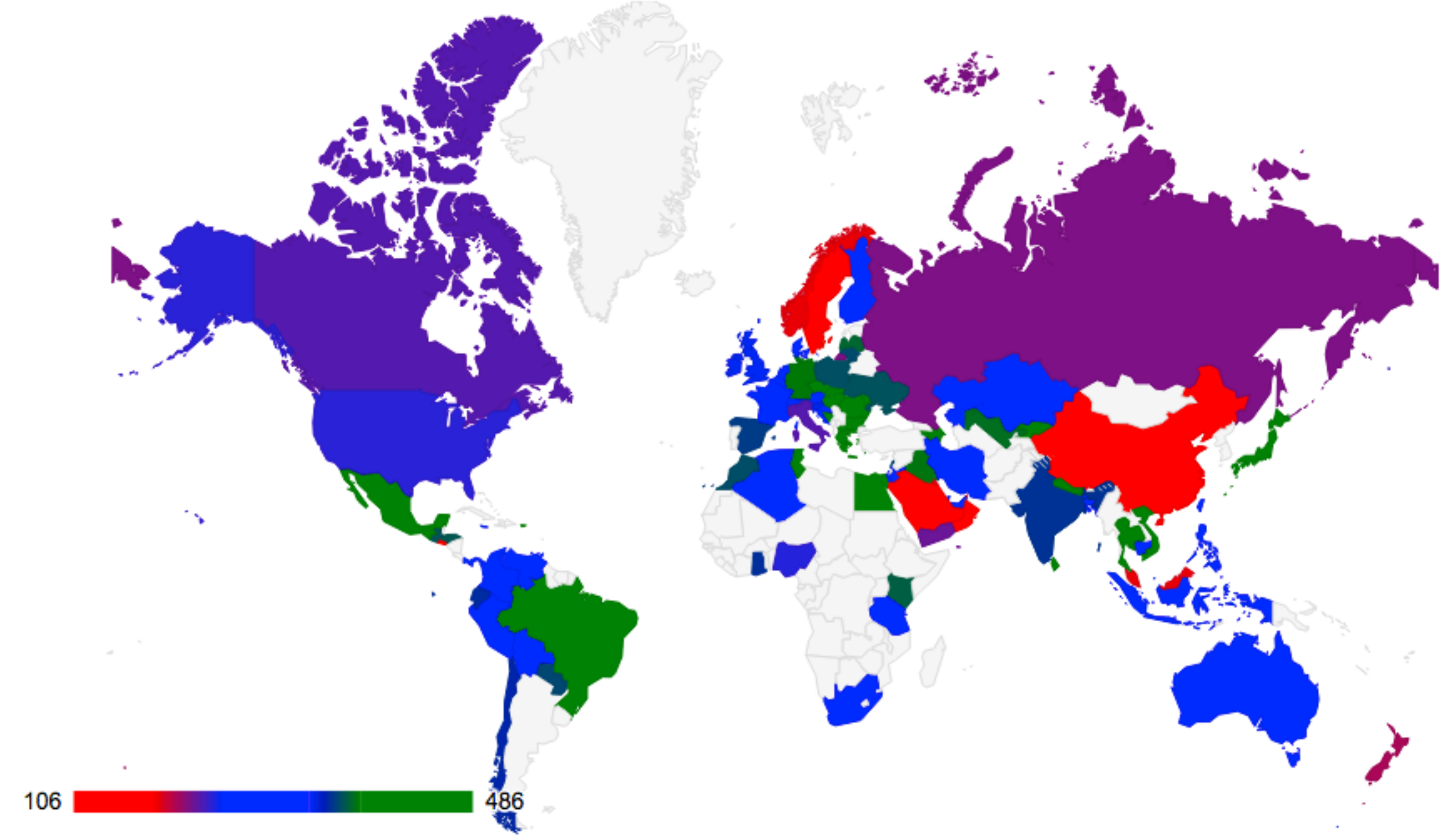}}
\subfloat[Browser-level Ad-Blockers] {\label{fig:browser_based_adb}
\includegraphics[width=0.49\columnwidth]{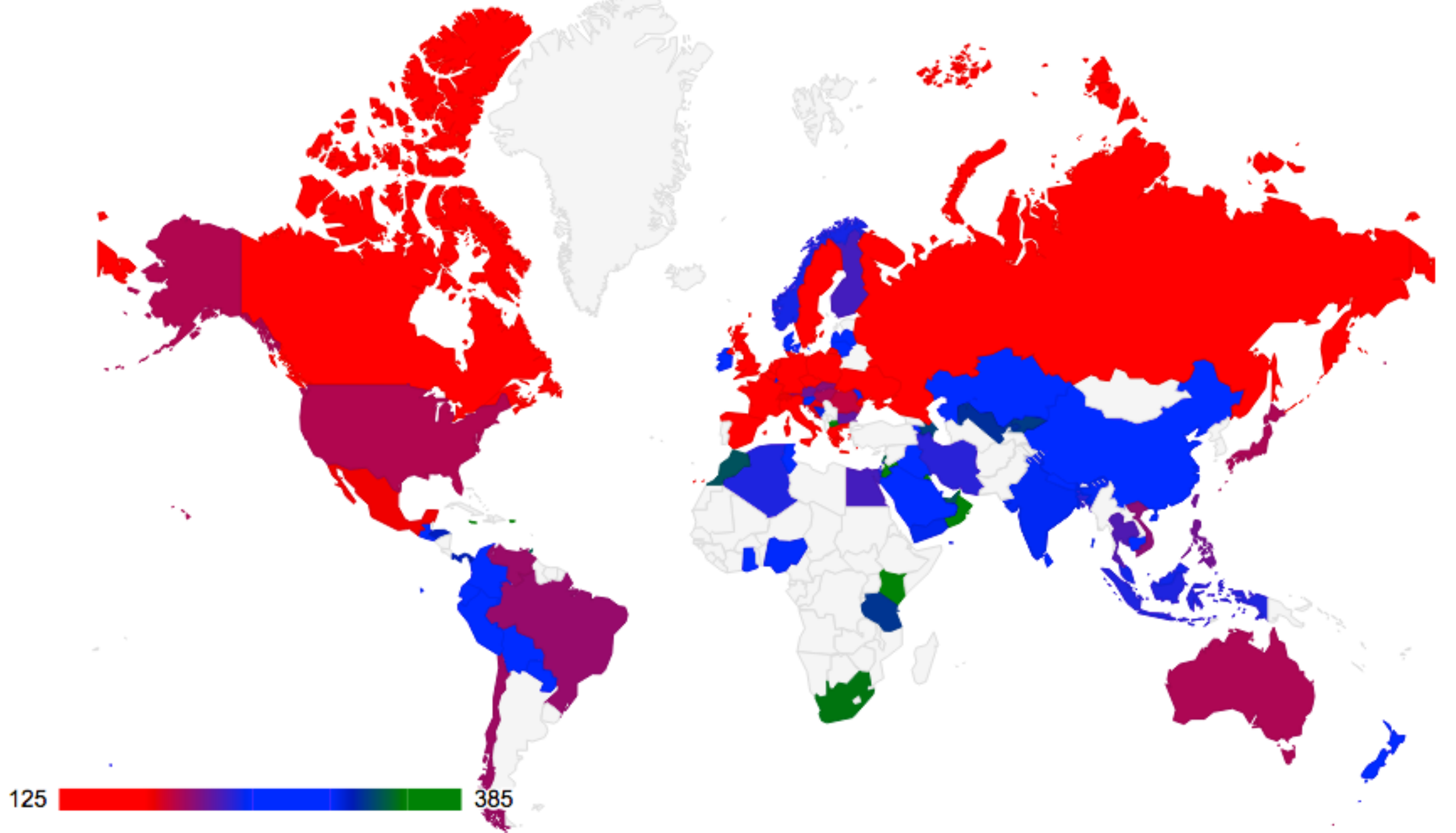}}\\

\caption{An overview of Ad-Blockers' popularity per country. The lower values of median rank, dark red colour, represent that Ad-Blockers are more popular in the  corresponding geographical region or country .}
\vspace{-0.40cm}
\label{fig:adblockers_geodistribution}
\end{figure}

The number of Ad-Blockers per country does not reveal the actual popularity of Ad-Blockers per country. We further analyse the popularity of each Ad-Blocker per country by measuring the median rank of all Ad-Blockers per country. Figure~\ref{fig:adblockers_geodistribution} shows the median rank of the analysed Ad-Blockers. Notably, VPN-based Ad-Blockers are more popular in countries such as China (median rank = 106), Macau (141), Saudi Arabia (152) and Oman (161) whereas browser-based Ad-Blockers are popular in France (125), Ukraine (132), and Germany(145). 
Compared to free Ad-Blockers, paid Ad-Blockers are popular in Taiwan (65), Spain (70), and Japan (70). 
%
When considering the median ranks of all Ad-Blockers per country, we observe that Ad-Blockers are popular in France (149), Russia (170), and Germany (176).

\section{Analysis of Ad-Blockers}
\label{sec:sanalysis}

In this section, we analyse the source code of the Ad-Blockers using static analysis. We examine the filter lists used by the Ad-Blockers and investigate the Ad-Blockers requesting sensitive permissions. Further, we evaluate the Ad-Blockers for the presence of third-party ads and tracking libraries, embedded in their code, and study their malware activities using VirusTotal. Finally, we present our analysis of Ad-Blockers' reviews on Google Play and report on users concerns about potential security and privacy issues as well as the inefficiency of Ad-Blockers.

\begin{table}[ht]
\centering
\scriptsize
\begin{tabular}{L{2.0cm} C{1.6cm}C{1.6cm}C{1.6cm}}
\toprule
& \multicolumn{2}{r}{}{{\bf Ad-Blockers}} & \\
\cline{2-4}
{\bf \# Filter Lists}  	&	{\bf Paid}  &	{\bf Free}  &	{\bf All}\\
\midrule
~~1	&	70\%		&	84\%		&	82\%		\\
~~2	&	30\%		&	8\%		&	10\%		\\
$\geq$3	&	0\%		&	8\%		&	8\%		\\

\bottomrule
\end{tabular}
\caption{Distribution of filter lists used by the analysed Ad-Blockers.}
\label{tab:filter_lists_in_classes}
\vspace{-0.4cm}
\end{table}

\begin{table}[ht]
\scriptsize
\centering
\tabcolsep=0.03cm
\begin{tabular}{L{2.5cm} C{0.6cm} C{0.6cm} C{0.6cm}L{4.0cm}}
\toprule
&\multicolumn{2}{r}{\bf Ad-Blockers} & \\
\cmidrule(r){2-4}
{\bf Filter List} & {\bf Paid} & {\bf Free} & {\bf All} & {\bf Description}\\  \midrule
CustomizedList &90\% & 39\% & 42\%& Customized list for ads/trackers\\
EasyList~\cite{easylist} &0\% & 18\% & 16\% & Filter list of trackers/ads\\
hpHosts~\cite{hpHost}  &10\% & 10\% & 10\% & Filter list of ad/tracker/malicious hosts\\
AcceptableAdsRules~\cite{exceptionrules} &0\% & 7\% & 7\%& List of non-intrusive, acceptable ads \\
MalwareHosts~\cite{malarehosts} & 0\% & 5\% & 5\% &Filter list of malicious hosts\\
FanboySocial~\cite{fanoby} & 0\% & 4\% & 4\%& Filter list of social buttons/widgets\\
AntiAdblockRules~\cite{antiadblcokfilters} & 0\% & 4\% & 4\%& Filters for evading anti-Adblock scripts \\
HostStevenBlack~\cite{sbhosts} & 0\% & 3\% & 3\%& Filter list of ad/tracker/malicious hosts\\
EasyPrivacy~\cite{easyprivacy} & 0\% & 3\% & 3\% &Filter list of trackers/analytics\\
AdSweep~\cite{adsweep}  &0\% & 3\% & 3\%&List of Rules to hide ads on websites \\
EasyListChina~\cite{EasyListChina} & 0\% & 2\% & 2\% & EasyList tailored for Chinese websites\\
YoyoHosts~\cite{yoyoadservers} & 0\% & 2\% & 2\% &Filter list of ad/tracker/malicious hosts\\
SomeOneCareHosts~\cite{someonecare} & 0\% & 1\% & 1\%&Filter list of ad/tracker/malicious hosts \\
MvpsHosts~\cite{mvpshosts}& 0\% & 1\% & 1\% & Blacklist of ad/tracker/malicious hosts\\
EasyListNEHide~\cite{easylistnoelemhide} & 0\% & 1\% & 1\% & EasyList without element hiding rules\\
AdAwayHosts~\cite{adaway}& 0\% & 1\% & 1\% & Filter list of mobile ad/tracker hosts\\

\bottomrule
\end{tabular}
\caption{Distribution of all filter lists in the analysed Ad-Blockers.}
\vspace{-0.4cm}
\label{tab:filterlist}
\end{table}

\subsection{Filter Lists in Use}
\label{sec:flists}

Ad-Blockers usually employ filter lists (e.g. Black lists) either downloaded locally or present on a remote server to block (or allow) third-party ads and tracking services. The decompilation of Ad-Blockers' APKs allows us to reveal the employed filter lists. Using {\tt Apktool}, we first decompile the Ad-Blockers' APKs and then search for specific keywords in the source code. The keywords include: ``easylists", ``abp", ``hosts", ``mvps", ``blacklist", ``blocklist", ``whitelist", ``exception rules", and ``acceptable ads". Finally, for each Ad-Blocker, we use {\tt JD-GUI} to manually inspect the decompiled source codes and to determine the employed filter lists. 

In Table~\ref{tab:filter_lists_in_classes}, we observe that 82\% of the analysed Ad-Blockers contain only one filter list. While 18\% of the Ad-Blockers use a composition of several filter lists.  

%
Table~\ref{tab:filterlist} shows the distribution of filter lists used by the Ad-Blockers. 90\% of the paid Ad-Blockers and 39\% of the free Ad-Blockers employ custom-built filter lists to block ads and tracking related traffic. We found that 16\% of the Ad-Blockers use EasyList~\cite{easylist} to filter advertisements on a browsed webpage. Our manual inspection of the code using {\tt ApkTool} and {\tt JD-GUI} reveals that all VPN-based Ad-Blockers (cf. Section~\ref{subsec:adblockingmechanisms}) such as F-Secure Freedome VPN and DashVPN use their custom-built filters lists to block ads and trackers. In particular, F-Secure Freedome VPN app blocks any traffic associated with web and mobile tracking including Google Ads, DoubleClick, and other popular tagging/analytics services such as Google Tag and ComScore~\cite{fsecureblacklist}. 
Moreover, 3\% of the Ad-Blockers such as Simple FLV and Fast Browser inject JavaScript codes, AdSweep~\cite{adsweep}, to hide {rather than block ads} on a browsed webpage. 

\subsection{Permission Analysis}
\label{sec:permissionanalysis}
We investigate how Ad-Blockers request Android permissions to access sensitive system resources. For each Ad-Blocker, we extract the requested permissions by parsing \emph{uses-permission} and \emph{service} tags in the \texttt{AndroidManifest.xml}. We exclude network-related permissions like Internet access which are inherent to Ad-Blockers. 

\begin{figure*}[ht]
\centering
\includegraphics[width=\textwidth]{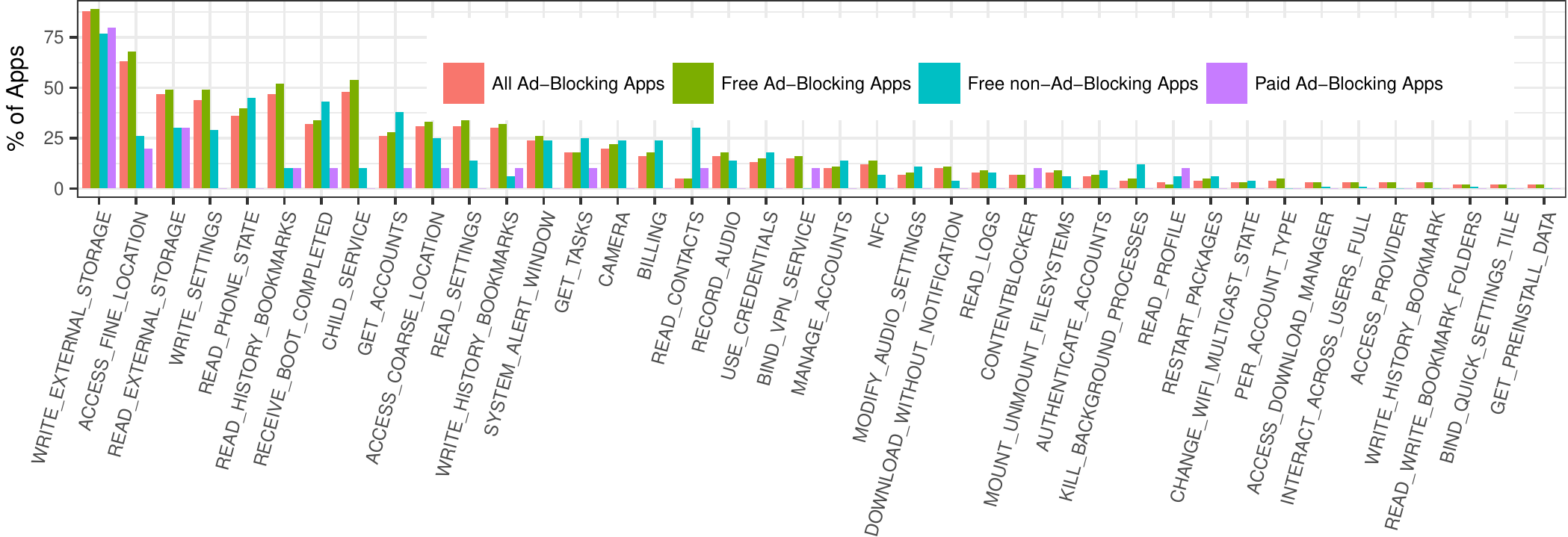}
\vspace{-0.5cm}
	\caption{Detailed comparison of Android permissions (x-axis) 
	requested by Ad-Blocking and free non-Ad-Blocking apps.}
    \label{fig:permissions_android}
    \vspace{-0.3cm}
\end{figure*}

Figure~\ref{fig:permissions_android} compares the permissions requested by Ad-Blockers with the baseline (cf. Dataset in Section~\ref{sec:adbchracterization}), which we included for reference. We use the method-to-permission mapping provided by Au~\etal~\cite{Au:2012:pscout} to investigate the 
sourcecode segments invoking the methods protected by each Android permission. Our analysis reveals that Ad-Blocking apps request access to permissions rarely 
requested by free non-Ad-Blocking apps.

Apps such as Anti-virus apps request the {\tt READ\_LOGS} permission to inspect other apps' activities~\cite{androidpermissions}\cite{Ikram_RSKP_16}. 
However, we observe that Ad-Blockers like UC Browser and DU Browser also request access to it. Android's documentation~\cite{androidpermissions} flags the {\tt READ\_LOGS} permission as highly sensitive as app developers may carelessly misuse Android's logging capabilities and (unintentionally) expose personal information (including passwords) to any other apps requesting it.

Several other permissions listed in Figure~\ref{fig:permissions_android} and detailed in Table~\ref{tab:identifierpermissions} appear unusual requirements for Ad-Blockers. For each case, we manually checked the legitimacy of these requests without finding a clear evidence of a deliberate abuse of the granted permissions. However, we found that spyware Ad-Blocking apps (which we further investigate in Section~\ref{sec:malwareanalysis}) request the {\tt READ\_SMS} permission to read text messages whereas AV apps may use it to scan text messages for possible malware presence. Similarly, apps requesting {\tt READ\_CONTACTS} incorporate functions in the likes of blocking text and calls from specific phone numbers or sharing features through SMS or email. 

\begin{table}[!ht]
\scriptsize
\centering
\tabcolsep=0.008cm
  \begin{tabular}{L{4.6cm} C{1.9cm} C{2.3cm}}
  	\toprule
   	 {\bf } &{\bf  All} & {\bf Free Non-} \\ 
  	
   	 {\bf Permission} &{\bf Ad-Blockers } & {\bf Ad-Blockers} \\ 
    	\midrule
		ACCESS\_FINE\_LOCATION & 63\% & 26\% \\
		READ\_HISTORY\_BOOKMARKS &47\% & 10\% \\
		GET\_ACCOUNTS & 26\% & 38\% \\
		SYSTEM\_ALERT\_WINDOW & 24\% & 24\% \\
		CAMERA & 24\% & 24\%\\
		RECORD\_AUDIO & 16\% & 14\%\\
		BIND\_VPN\_SERVICE & 14\% & 0\%\\
		DOWNLOAD\_WITHOUT\_NOTIFICATION & 10\% & 4\% \\
		READ\_LOGS & 8\% & 2\% \\
		READ\_CONTACTS & 5\% & 30\% \\
		READ\_CALL\_LOG & 2\% & 13\% \\
		READ\_SMS & 1\% & 16\% \\
		READ\_CONTACT & 1\% & 16\% \\
		READ\_CALENDAR & 1\% & 9\% \\
		

		\bottomrule
	 \end{tabular}
	\caption{Privacy sensitive permissions requested by Ad-Blockers.}
	\label{tab:identifierpermissions}
	\vspace{-0.45cm}
\end{table}

\subsection{Embedded Ads and Tracking Libraries}
\label{subsec:eatl}
We examine the presence of embedded libraries for tracking or advertising purposes in the source-code of each Ad-Blocker. 
In order to conduct our analysis, we use {\tt ApkTool} to decompile each Ad-Blocker 
and perform a dictionary-based search for ads and tracking libraries inside the decompiled source code. We rely on \cite{seneviratne2015measurement} and compile a comprehensive dictionary of 338 mobile tracking libraries. 

\begin{figure*}[ht]
\centering
    \includegraphics[width=\textwidth]{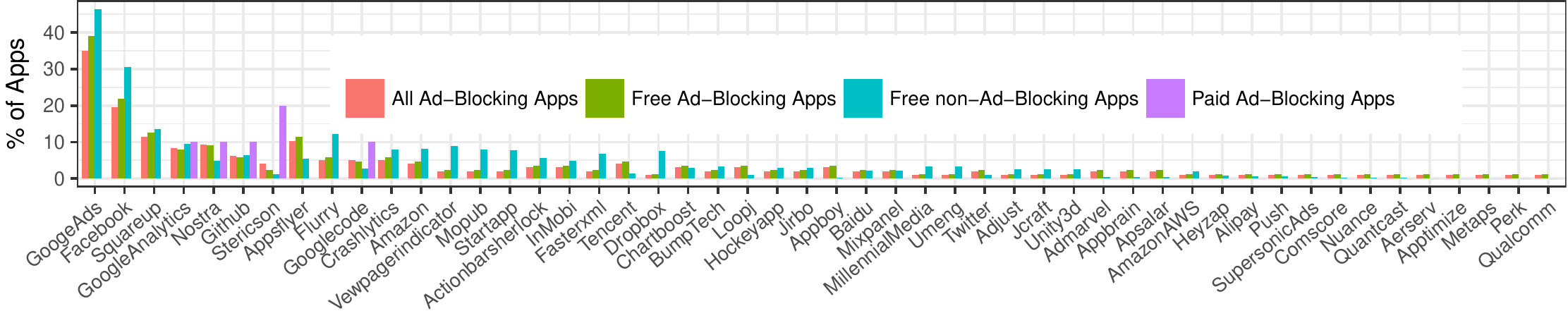}
    \vspace{-0.5cm}
    \caption{Third-party libraries (x-axis) in Ad-Blocking and free non-Ad-Blocking apps. 
		}
	\label{fig:trackers_in_histogram}
	\vspace{-0.4cm}
\end{figure*}

\begin{table}
\centering
\scriptsize
\tabcolsep=.35cm
\begin{tabular}{lcccc}
\toprule
& \multicolumn{2}{r}{}{{\bf Ad-Blockers}} & & {\bf Free} \\
\cline{2-4}
{\bf \# Trackers}  	&	{\bf Paid}  &	{\bf Free}  &	{\bf All} &	{\bf non-Ad-Blocking Apps}\\
\midrule
~~~0	&	40\%		&	31\%		&	32\%		&	16\%		\\
~~~1	&	50\%		&	14\%		&	18\%		&	7\%		\\
~~~2	&	10\%		&	13\%		&	12\%		&	13\%		\\
~~~3	&	0\%		&	10\%		&	9\%		&	16\%		\\
~~~4	&	0\%		&	11\%		&	10\%		&	14\%		\\
$\geq$5	&	0\%	&	21\%		&	19\%		&	34\%		\\
\bottomrule
\end{tabular}
\vspace{-0.2cm}
\caption{Analysis of third party ads and tracking libraries in Ad-Blockers and free non-Ad-Blocking apps.}
\label{tab:trackeranalysis_vpn_classes}
\vspace{-0.4cm}
\end{table}

Table~\ref{tab:trackeranalysis_vpn_classes} compares the number of ads and tracking libraries used 
by Ad-Blocking apps with the presence of ads and tracking libraries in the reference set of 500 
free non-Ad-Blocking apps. We observe that 68\% of the Ad-Blocking apps embed at least one third-party ads and tracking library in their
code. The penetration of tracking libraries in Ad-Blocking apps is however significantly lower than in the reference set of 500 non-Ad-Blocking apps with 
84\% of the latter having at least one embedded tracking library. 

Since Ad-Blockers intend to block trackers and (intrusive) advertisements, the lower presence of tracking and advertisement libraries is actually meaningful. Nevertheless, we identified at least one targeted ads library in 68\% of the Ad-Blockers claiming (as mentioned in their apps description on Google Play) to block advertisements. While paid apps are often thought to be free from ads (as their business model is supposedly not driven by advertising), we observed a disturbing 60\% of the paid Ad-Blockers having at least one embedded third-party ad and tracking library. 
In particular, Photon Flash Player \& Browser and Perk Browser--two popular apps, which combined have more than 11M installs-- have the highest number of embedded third-party tracking libraries: 13 and 11 tracking libraries respectively. In general, 19\% of the Ad-Blockers have at least five third-party ads and tracking libraries.

Figure~\ref{fig:trackers_in_histogram} ranks the trackers in all 
analysed Ad-Blocking apps. Google Ads and Facebook social analytics are the most popular ones 
among our corpus of Ad-Blockers. A closer examination at the long-tail of the distribution reveals that the least popular third-party tracking libraries are more common in Ad-Blockers 
than in free non-Ad-Blocking apps. For instance, Ad-Blockers like Opera Browser
and DU Browser, each has over 10M installs, integrate libraries like Appboy\footnote{https://www.appboy.com} and Tencent\footnote{http://tencent.com/en-us/ps/adservice.shtml} for tracking and delivering targeted ads.

\subsection{Malware Analysis}
\label{sec:malwareanalysis}

We explore the presence of malware in the analysed Ad-Blockers. 
In order to effectively identify any malware activity on mobile apps, it is critical to rely on multiple malware scanning tools
as malicious components may be designed to circumvent some AV tools and malware scanners. 
To improve the confidence of our malware scans, we rely on VirusTotal, an online service that aggregates the scanning capabilities provided by multiple AV
tools.

\begin{table}[ht]
\small
\scriptsize
\centering
\tabcolsep=0.08cm
\begin{tabular}{C{0.3cm} L{2.2cm}C{1cm}C{1cm}C{1cm}C{1cm}C{1cm}}
\toprule
{\bf \#}  & {\bf App ID} &{\bf Price}&  {\bf Rating} & {\bf Installs} & {\bf AV-rank} & {\bf DevLoc} \\
\midrule
1 & Deep Search Browser&Free & 2.6 & 1K& 10 & PK\\
2&	sFly Network Booster&Free	&4.3&1K	&	10& CN	\\
3& Faster Browser Ever&Free & 4.0 & 1K & 6 & RU\\ 
4 & FastCat&Free & 4.5 & 100 & 6& AE\\
5& Magneto Browser&Free & 4.0 & 100 & 4& IN \\ 
6& Adskip Browser&Free & 2.0 & 5K & 4& CN \\ 
7&	Maxthon Browser&Free	 	&4.4 &10M	 & 3	& CN\\

\bottomrule
\end{tabular}
\vspace{-0.20cm}
\caption{Ad-Blockers with a VirusTotal AV-rank $\geq$ 3. DevLoc represents the geolocation of Ad-Blockers' developers.}
\label{tab:malwares}
\vspace{-0.6cm}
\end{table}

We automate our malware scanning process by using VirusTotal's public API. 
After completing the scan, VirusTotal generates a report that  
indicates which of the participating AV scanning tools detected 
any malware activity in the app and the corresponding malware signature (if any). 
Given that a scanning tool may produce false positives, 
we rely on the ``AV-rank'' metric (\ie the
number of affiliated AV tools that identified any malware activity) 
to reason about the maliciousness of an Ad-Blocker.
We consider an ``AV-rank'' threshold $\geq$ 3 as a signal for malware presence in Ad-Blockers. Additionally, we glean developers' metadata from Google Play and AppAnnie to investigate the geolocations of developers of malicious Ad-Blockers.

Table~\ref{tab:malwares} lists 
the Ad-Blocking apps ranked by their AV-rank. 
We include developers' geolocation and apps' Google Play rating and the number of installs for each 
app for reference. 
21\% of the analysed Ad-Blocking apps have at least one
positive malware report according to VirusTotal with 
7\% of the Ad-Blockers have an ``AV-rank'' above our threshold. 
The malware signatures correspond to five different classes of malware: 
Adware (22\%), Trojan (23\%), Malvertising (16\%), and Riskware (39\%).  
We observe that malware developers are located in Russia and Asian countries such as India, China, Pakistan, and United Arab Emirates. 

Next, we analyse the permission-protected API access of the malicious Ad-Blockers. Maxthon Browser 
incorporates Adware on its source-code and requests the intrusive {\tt SYSTEM\_ALERT\_WINDOW} permission which allows the Maxthon Browser to draw window alerts (e.g., full screen Ad windows) on top of any other active app. 
Likewise, DU Browser, which incorporates Riskware elements according to VirusTotal,
requires the {\tt READ\_LOGS}, {\tt READ\_PHONE\_STATE}, {\tt READ\_HISTORY\_BOOKMARKS}, {\tt READ\_SETTINGS}, and {\tt WRITE\_SETTINGS} permissions to read users' settings, hijack bookmarks, change browser's start page, and link web searches to potentially lower ranked sites.

\subsection{Apps Reviews Analysis}
\label{sec:userperception}

We use negative users comments to capture the perceptions and concerns about the Ad-Blocking functionality of the analysed Ad-Blockers. Our reasoning to focus our analysis on negative reviews, 1- and 2-star reviews appeared on Google Play, for popular apps is that any serious Ad-Blocking-related concern exposed by a user should receive a negative review.

\begin{table}[t!]
\scriptsize
\tabcolsep=0.03cm
\centering
\begin{tabular}{L{3.5cm} C{5cm}}
\toprule
{\bf Complaint Category}   & {\bf \% of negative reviews ($N=\totalReviewsChecked$)} \\
\midrule
1 Star Reviews & 13,764 \\ 
2 Star Reviews & 6,244 \\
 \hline
Allowing/not-blocking Ads & 16\%\\
Bugs \& battery life & 7\% \\
Abusive permissions & 1.5\% \\ 
Malware/fraud reports & 0.65\% \\
\bottomrule
\end{tabular}
\vspace{-0.1cm}
\caption{Classification of negative reviews for Ad-Blockers in Google Play.}
\label{tab:adblockers_userreviews_classification}
\vspace{-0.55cm}
\end{table}

To better identify whether users publicly report any Ad-Blocking
concerns after using each Ad-Blocker, we analyse (with manual supervision) 
the content of \totalReviewsChecked negative  
reviews for 88\% of the analysed Ad-Blockers\footnote{12\% of the Ad-Blockers do not have negative reviews or do not have reviews at all.}. We label the app reviews 
into the 4 categories listed in Table~\ref{tab:adblockers_userreviews_classification}
which cover from performance concerns and bugs to
different types of Ad-Blocking concerns as well as abusive or intrusive permission requests. 
We exclude from our analysis any reviews related with usability concerns such as bugs and crashes. 
Note that 7\% of the complaints are about bugs, crashes and other performance aspects
such as bugs and battery-life overhead. 

\begin{figure}[ht]
\centering
\subfloat[]{\label{fig:dashnetsettings} 
\includegraphics[width=0.49\columnwidth, height=5.0cm,keepaspectratio]{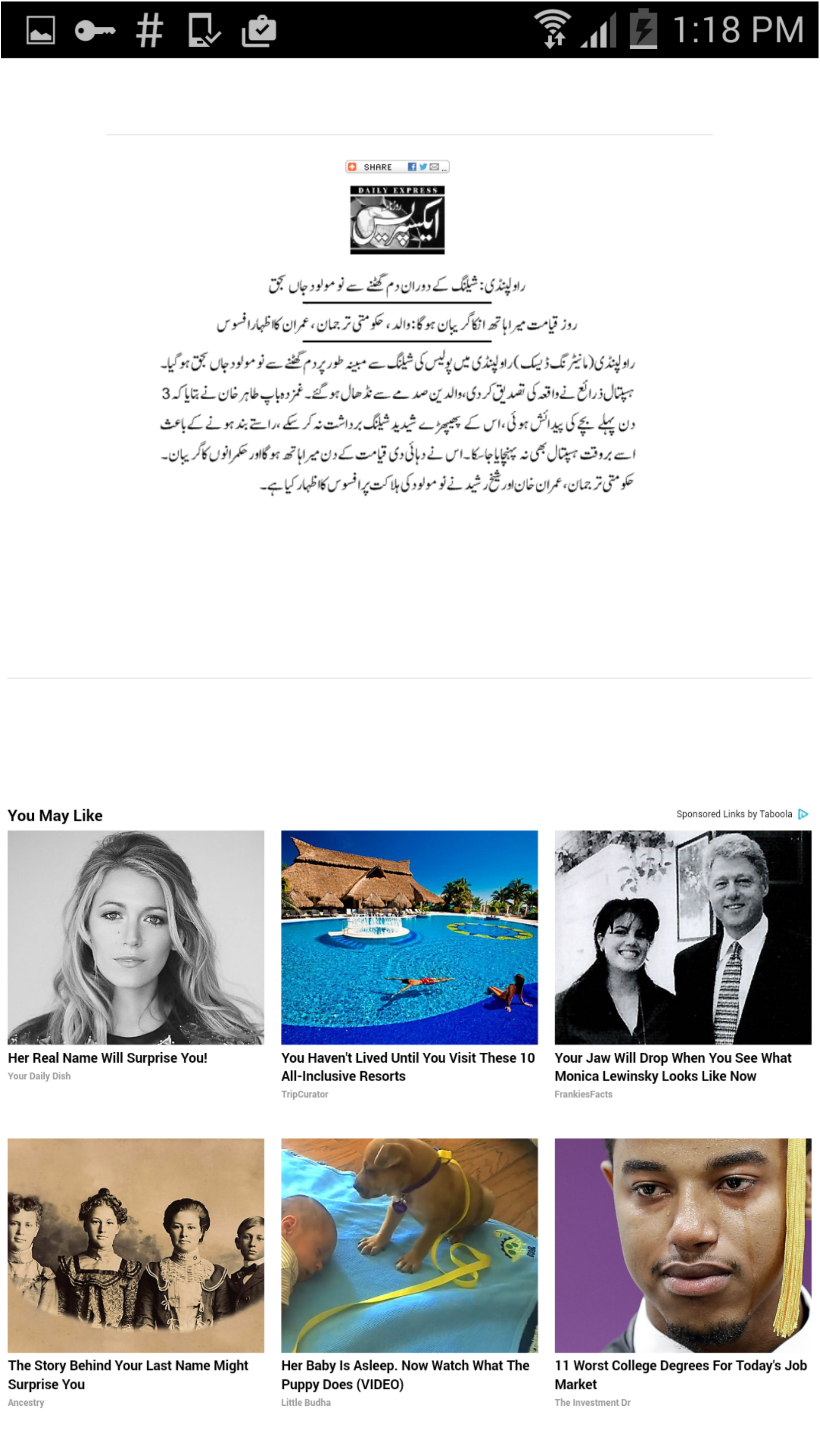}}
\subfloat[] {\label{fig:dashnetads} \hspace*{1.9em}
\includegraphics[width=0.49\columnwidth, height=5.0cm, keepaspectratio]{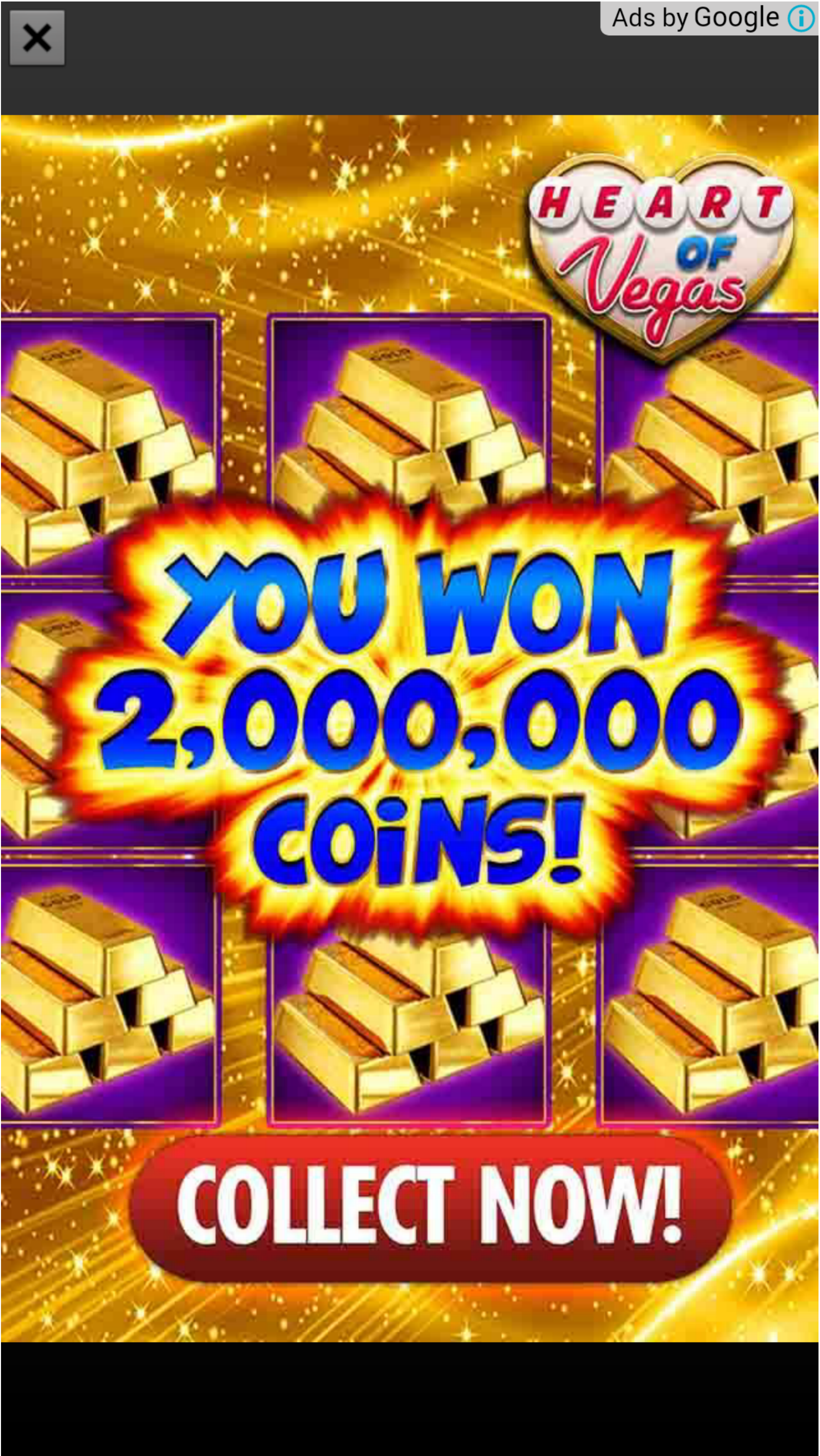}}
\vspace{-0.2cm}
\caption{Screenshots showing inefficiency of DashVPN, a system-level Ad-Blocker: \textcolor{red}{(a)}  DashVPN does not block click-baits from Taboola (\texttt{taboola.com}) and \textcolor{red}{(b)} shows full-screen ads from Google.}
\vspace{-0.65cm}
\label{fig:inefficiencyofAdblockers}
\end{figure}

We observe that 16\% of the negative reviews report on the allowing or displaying ads in In-App Ad-banner or visited websites (cf. Figure~\ref{fig:androidappmodel}). We found that 77\% of the analysed Ad-Blockers have at least one negative review about their inefficiency to block ads. We manually check the negatively reviewed apps and found that 24\% of the apps show ads and display full-screen advertisements (e.g., Figure~\ref{fig:inefficiencyofAdblockers}). 

Notably, 1.5\% of the negative reviews report on the intrusiveness and sensitive permissions requests of 37\% of the analysed Ad-Blockers. We found that these 37\% Ad-Blockers have at least one third-party ads and tracking library embedded in their code (as per the analysis in Section~\ref{subsec:eatl}). Moreover, we observed that 130 ($ \simeq$ 0.65\%) of the reviews explicitly report malware or fraudulent activity in 22\% of the analysed Ad-Blockers, listed in Table~\ref{tab:complaints}. Six of the apps reported as malicious by end users are also considered as malicious by VirusTotal (cf. Section~\ref{sec:malwareanalysis}). Interestingly however, and despite the presence of some awareness and public comments about the potential security and privacy issues of Ad-Blockers, these apps still exhibit relatively high average ratings and a significant number of installs (cf. Table~\ref{tab:complaints}). Our analysis suggests that a significant number of Ad-Blockers do not fulfil their main ``intended'' Ad-Blocking functionality. 

\begin{table}[ht]
\scriptsize
\centering
\tabcolsep=0.03cm
\begin{tabular}{L{0.3cm}L{3.4cm}C{0.7cm}C{1.1cm}C{0.8cm}C{0.8cm}C{1.4cm} } 
\toprule
{\bf \#} & {\bf App}  	& {\bf Class } & {\bf NR-Ratio} & {\bf Rating} & {\bf Installs}  & {\bf AV-positive} \\ 
 
\midrule
1 & UC Browser - Fast Download	&Free&	32\%	&	4.5	&	100M	&	\cmark	\\
2 &Opera browser - latest news	&Free&	15\%	&	4.3	&	100M	&		\\
3 &Yandex Browser for Android	&Free&	12\%	&	4.5	&	10M	&	\cmark	\\
4 &Maxthon Browser - Fast\&Secure	&Free&	21\%	&	4.4	&	10M	&	\cmark	\\
5 &Photon Flash Player \& Browser	&Free&	28\%&	3.7	&	10M	&		\\
6 &CM Browser - Fast \& Light	&Free&	15\%&	4.5	&	50M	&		\\
7 &Dolphin - Best Web Browser	&Free&	25\%	&	4.5	&	50M	&		\\
8 &Adblock Browser for Android	&Free&	26\%	&	4.1	&	5M	&	\cmark	\\
9 &APUS Browser - Fast Download	&Free&	22\%	&	4.5	&	5M	&	\cmark	\\
10 &Free Adblocker Browser	&Free&	8\%	&	4.4	&	1M	&		\\
11 &Dolphin Jetpack - Fast \& Flash	& Free& 31\%		&	4.1	&	1M	&		\\
12 &F-Secure Freedome VPN	&	Free	&11\%	&	4.3	&	1M	&		\\
13 &Dolphin Zero Incognito Browser	&	Free &	17\%	&	4.2	&	1M	&		\\
14 &Adblock Plus (Samsung Browser)	&	Free & 37\%	&	3.9	&	500K	&		\\
15 &Mercury Browser for Android	&	Free & 19\%	&	4.3	&	500K	&		\\
16 &Ghostery Privacy Browser	&	Free & 	21\%	&	4.1	&	500K	&		\\
17 &Rocket Browser	&	Free	&	18\%	&	4.4	&	100K	&		\\
18 &AC Browser	&	Free & 	6\%	&	4.5	&	100K	&		\\
19 &NetGuard - no-root firewall	&	Free &	19\% 	&	4.2	&	100K	&		\\
20 &Dee Browser	&	Free & 9\%	&	4.1	&	50K	&	\cmark	\\
\bottomrule
\end{tabular}
\vspace{-0.10cm}
\caption{List of Ad-Blockers that are considered as malicious by users 
	in Google Play reviews and by VirusTotal (AV-positive column). For each Ad-Blocker, the \emph{NR-Ratio} represents the ratio of the number of negative users' comments to the total number of all users' comments.}
\vspace{-0.2cm}
\label{tab:complaints} 
\end{table}

\section{Related work}
\label{sec:rwork}

The Web has evolved, with increase in the prevalence and complexity of tracking mechanisms since 1996, into a tangled mass of third-party tracking domains embedded into first-party webpages~\cite{197133}. Research studies have reveal that {the top 5\% of webpages} embed 100 third party domains~\cite{wills:2016:}. Among other services such as provision of multimedia services via content delivery networks and user-interactions, these third parties provide a variety of services such as tracking users, serving ads, and performing site analytics. Notably, the Alexa's top 10K websites have an average 34 third-party tracking and advertisement services~\cite{sEnglehardt2016}.

Several studies highlighted the privacy risks associated with Android apps over-requesting Android permissions for third-party tracking, advertising and analytic services~\cite{seneviratne2015measurement} using techniques like static analysis~\cite{Au:2012:pscout}, taint analysis~\cite{enck2014taintdroid}, and OS modifications~\cite{Jeon:spsm2012}. Previous research also adapted techniques for malware detection such as signature analysis~\cite{bose2008behavioral} to the mobile context in order to identify potential malicious activity of mobile apps. Using static code and dynamic analysis techniques, Ikram et al.,~\cite{Ikram_RSKP_16} measured mobile VPN apps and identified several security and privacy issues in 283 different VPN permission-enabled Android apps. The authors also highlighted an alarming mismatch in apps' descriptions on Google Play and their actual functionalities. 


Studies have examined the effectiveness of web Ad-Blockers~\cite{Wang:2015}~\cite{ikram2016towards}. Ikram et al.,~\cite{ikram2016towards} evaluated the effectiveness of five different web Ad-Blocking plugins and proposed an improved machine-learning based solution to strike the balance between blocking tracking/advertising domains and allowing domains that serve useful content such as CDNs. Wills and Uzunoglu~\cite{wills:2016:}, investigated the default and fully configured settings of Ad-Blocking plugins. They observed that the default as well as the fully configured settings, with composite filter lists, of the plugins are inefficient to block ads- and tracking-related traffic. 
In contrast to the previous work, this paper present the first characterisation study of mobile Ad-Blocking apps with a focus on security and privacy offered by these apps. 

\section{Conclusion and Future Work}
\label{sec:conclusion}

The increasing number of mobile Ad-Blockers available on apps' markets such as Google Play and the growing number of complaints raised by users indicate serious ineffectiveness or usability issues thus necessitate the urge to analyse this unexplored eco-system. The average mobile user rates Ad-Blockers positively even when they have malware presence. According to our study, 16\% of negative reviews are {related to (or concerned with)} the ineffectiveness of the Ad-Blockers suggesting serious performance issues. Moreover, our analysis of Ad-Blockers, reviewed by users, reveals that several Ad-Blockers such as F-Secure Freedome VPN caused several usability issues while running other installed apps or surfing the web. 
We believe that our work could be extended to study {the ineffectiveness of the Ad-Blockers}. As a future work, we plan to complement the insights provided by our analysis with a comprehensive set of active tests to reveal the runtime behavioural aspects of the Ad-Blockers. {In essence, we aim to device a testbed to analyse the Ad-Blockers' in-effectiveness, to validate their data- and energy-saving claims, and to evaluate their resilience against anti-Ad-Blocking JavaScripts on the Web.} 

\bibliographystyle{abbrv}
\bibliography{sigproc}

\begin{thebibliography}{10}

\bibitem{antiadblcokfilters}
{AakList (Anti-Adblock Killer)}.
\newblock
  {https://github.com/reek/anti-adblock-killer/blob/master/anti-adblock-killer-filters.txt}.

\bibitem{adaway}
{AdAway Hosts}.
\newblock {https://adaway.org/hosts.txt}.

\bibitem{abpplugin}
{Adblock Plus Plugin}.
\newblock {https://adblockplus.org}.

\bibitem{adsweep}
{AdSweep}.
\newblock {https://web.archive.org/web/
  20121126071410/http://www.adsweep.org/}.

\bibitem{exceptionrules}
{Allow non-intrusive advertising}.
\newblock {https://easylist-downloads.adblockplus.org/exceptionrules.txt}.

\bibitem{androidpermissions}
{Android Permissions}.
\newblock {http://developer.android.com/guide/topics
  /security/permissions.html}.

\bibitem{appannie}
{App Annie Insights}.
\newblock https://www.appannie.com/en/insights/.

\bibitem{EasyListChina}
{China+EasyList}.
\newblock
  {https://easylist-downloads.adblockplus.org/easylistchina+easylist.txt}.

\bibitem{easylist}
{EasyList}.
\newblock {https://easylist.to/easylist/easylist.txt}.

\bibitem{easylistnoelemhide}
{EasyList Without Element Hiding Rules}.
\newblock https://easylist-downloads.adblockplus.org/easylist\_noelemhide.txt.

\bibitem{easyprivacy}
{EasyPrivacy}.
\newblock {https://easylist.to/easylist/easyprivacy.txt}.

\bibitem{fsecureblacklist}
{F-Secure Freedome Anti-Tracking Feature Explained}.
\newblock
  {https://community.f-secure.com/t5/F-Secure/F-Secure-Freedome-Anti-Tracking/ta-p/52153}.

\bibitem{fanoby}
{Fanboy's Social Blocking List}.
\newblock {https://easylist-downloads.adblockplus.org/fanboy-social.txt}.

\bibitem{unofficialg}
{Google Play Unofficial Python API}.
\newblock {https://github.com/egirault/googleplay-api}.

\bibitem{hpHost}
{hpHosts}.
\newblock {http://www.hosts-file.net}.

\bibitem{malarehosts}
{Malware Domain Hosts Lists}.
\newblock {https://www.malwaredomainlist.com/hostslist/hosts.txt}.

\bibitem{mvpshosts}
{MVPS Hosts Lists}.
\newblock {http://winhelp2002.mvps.org/hosts.txt}.

\bibitem{raccoon}
{Raccon APK Downloader}.
\newblock {http://www.onyxbits.de/raccoon}.

\bibitem{someonecare}
{SomeOneCare Hosts Lists}.
\newblock {http://someonewhocares.org/hosts/ hosts}.

\bibitem{sbhosts}
{Steven Black Hosts}.
\newblock {https://github.com/StevenBlack/hosts}.

\bibitem{yoyoadservers}
{YoYo Ad Server List}.
\newblock {https://pgl.yoyo.org/adservers/}.

\bibitem{Au:2012:pscout}
K.~W.~Y. Au, Y.~F. Zhou, Z.~Huang, and D.~Lie.
\newblock {PScout: Analyzing the Android Permission Specification}.
\newblock In {\em CCS}, 2012.

\bibitem{bose2008behavioral}
A.~Bose, X.~Hu, K.~G. Shin, and T.~Park.
\newblock {Behavioral Detection of Malware on Mobile Handsets}.
\newblock In {\em {ACM MobiSys}}, 2008.

\bibitem{enck2014taintdroid}
W.~Enck, P.~Gilbert, B.-G. Chun, L.~P. Cox, J.~Jung, P.~McDaniel, and A.~N.
  Sheth.
\newblock {TaintDroid: An Information Flow Tracking System for Real-Time
  Privacy Monitoring on Smartphones}.
\newblock {\em CACM}, 2014.

\bibitem{sEnglehardt2016}
S.~Englehardt and A.~Narayanan.
\newblock {Online Tracking: A 1-million-site Measurement and Analysis }.
\newblock In {\em {CCS}}, 2016.

\bibitem{ikram2016towards}
M.~Ikram, H.~J. Asghar, M.~A. Kaafar, B.~Krishnamurthy, and A.~Mahanti.
\newblock Towards seamless tracking-free web: Improved detection of trackers
  via one-class learning.
\newblock {\em {PETS}}, 2017.

\bibitem{Ikram_RSKP_16}
M.~Ikram, N.~V. Rodriguez, S.~Seneviratne, D.~Kaafar, and V.~Paxson.
\newblock An analysis of the privacy and security risks of android {VPN}
  permission-enabled apps.
\newblock In {\em ACM IMC}, 2016.

\bibitem{Jeon:spsm2012}
J.~Jeon, K.~K. Micinski, J.~A. Vaughan, A.~Fogel, N.~Reddy, J.~S. Foster, and
  T.~Millstein.
\newblock {Dr. Android and Mr. Hide: Fine-grained Permissions in Android
  Applications}.
\newblock In {\em ACM SPSM}, 2012.

\bibitem{197133}
A.~Lerner, A.~K. Simpson, T.~Kohno, and F.~Roesner.
\newblock Internet jones and the raiders of the lost trackers: An
  archaeological study of web tracking from 1996 to 2016.
\newblock In {\em {USENIX Sec.}}, 2016.

\bibitem{seneviratne2015measurement}
S.~Seneviratne, H.~Kolamunna, and A.~Seneviratne.
\newblock {A Measurement Study of Tracking in Paid Mobile Applications}.
\newblock In {\em ACM WiSec}, 2015.

\bibitem{Wang:2015}
N.~Wang, B.~Zhang, B.~Liu, and H.~Jin.
\newblock Mobilehci.
\newblock 2015.

\bibitem{wills:2016:}
C.~E. Wills and D.~C. Uzunoglu.
\newblock What ad blockers are (and are not) doing.
\newblock In {\em HotWeb}, 2016.

\end{thebibliography}

\onecolumn
\scriptsize
\begin{longtable}{lllccccc}
\toprule
{\bf \#} & {\bf Title}	&	{\bf App ID}	&{\bf Category}	&{\bf Rating}	&{\bf Price}	&{\bf Class}	&{\bf Installs}	\\
\midrule
1	&	Opera browser - news \& search	&	com.opera.browser	&	Comm. 	&	4.34	&	Free	&	Browser	&	100M	\\
2	&	Opera Mini - fast web browser	&	com.opera.mini.native	&	Comm. 	&	4.37	&	Free	&	Browser	&	100M	\\
3	&	UC Browser - Fast Download	&	com.UCMobile.intl	&	Comm. 	&	4.49	&	Free	&	Browser	&	100M	\\
4	&	Firefox Browser fast \& private	&	org.mozilla.firefox	&	Comm. 	&	4.38	&	Free	&	Browser	&	100M	\\
5	&	CM Browser - Fast \& Light	&	com.ksmobile.cb	&	Comm. 	&	4.54	&	Free	&	Browser	&	50M	\\
6	&	Dolphin - Best Web Browser	&	mobi.mgeek.TunnyBrowser	&	Comm. 	&	4.53	&	Free	&	Browser	&	50M	\\
7	&	Photon Flash Player \& Browser	&	com.appsverse.photon	&	Comm. 	&	3.73	&	Free	&	Browser	&	10M	\\
8	&	DU Browser—Browse fast \& fun	&	com.baidu.browser.inter	&	Comm. 	&	4.32	&	Free	&	Browser	&	10M	\\
9	&	Dolphin Browser Express: News	&	com.dolphin.browser.express.web	&	Comm. 	&	4.1	&	Free	&	Browser	&	10M	\\
10	&	Maxthon Web Browser	&	com.mx.browser	&	Comm. 	&	4.44	&	Free	&	Browser	&	10M	\\
11	&	Yandex Browser for Android	&	com.yandex.browser	&	Comm. 	&	4.47	&	Free	&	Browser	&	10M	\\
12	&	Best VPN - Free Unlimited VPN	&	com.northghost.touchvpn	&	Tools	&	4.39	&	Free	&	VPN	&	5M	\\
13	&	APUS Browser - Fast Download	&	com.apusapps.browser	&	Person.	&	4.48	&	Free	&	Browser	&	5M	\\
14	&	Opera browser beta	&	com.opera.browser.beta	&	Comm. 	&	4.25	&	Free	&	Browser	&	5M	\\
15	&	Adblock Browser for Android	&	org.adblockplus.browser	&	Comm. 	&	4.08	&	Free	&	Browser	&	5M	\\
16	&	F-Secure Freedome VPN	&	com.fsecure.freedome.vpn.security.privacy.android	&	Tools	&	4.29	&	Free	&	VPN	&	1M	\\
17	&	Opera Free VPN - Unlimited VPN	&	com.opera.vpn	&	Tools	&	4.35	&	Free	&	VPN	&	1M	\\
18	&	Dolphin Browser (JP)	&	com.dolphin.browser.android.jp	&	Comm. 	&	4.15	&	Free	&	Browser	&	1M	\\
19	&	Dolphin Jetpack - Fast \& Flash	&	com.dolphin.browser.engine	&	Comm. 	&	4.15	&	Free	&	Browser	&	1M	\\
20	&	Dolphin Zero Incognito Browser	&	com.dolphin.browser.zero	&	Social	&	4.2	&	Free	&	Browser	&	1M	\\
21	&	Free Adblocker Browser	&	com.hsv.freeadblockerbrowser	&	Comm. 	&	4.39	&	Free	&	Browser	&	1M	\\
22	&	Speed Booster for Android 	&	mobi.mgeek.browserfaster	&	Social	&	4.17	&	Free	&	Browser	&	1M	\\
23	&	Lightning Web Browser	&	acr.browser.barebones	&	Comm. 	&	4.15	&	Free	&	Browser	&	500K	\\
24	&	Ghostery Privacy Browser	&	com.ghostery.android.ghostery	&	Comm. 	&	4.13	&	Free	&	Browser	&	500K	\\
25	&	Mercury Browser for Android	&	com.ilegendsoft.mercury	&	Prod.	&	4.29	&	Free	&	Browser	&	500K	\\
26	&	Link Bubble	&	com.linkbubble.playstore	&	Person.	&	4.18	&	Free	&	Browser	&	500K	\\
27	&	Sleipnir Mobile - Web Browser	&	jp.co.fenrir.android.sleipnir	&	Comm. 	&	3.89	&	Free	&	Browser	&	500K	\\
28	&	Adblock Plus (Samsung Browser)	&	org.adblockplus.adblockplussbrowser	&	Comm. 	&	3.89	&	Free	&	Browser	&	500K	\\
29	&	Dash VPN	&	com.actmobile.dashvpn	&	Busin.	&	4.03	&	Free	&	VPN	&	100K	\\
30	&	NetGuard - no-root firewall	&	eu.faircode.netguard	&	Tools	&	4.24	&	Free	&	VPN	&	100K	\\
31	&	Adblock Browser	&	adblock.browser.lightning	&	Prod.	&	3.73	&	Free	&	Browser	&	100K	\\
32	&	Emo Ads Blocker Browser	&	ads.blocker.browser	&	Tools	&	3.38	&	Free	&	Browser	&	100K	\\
33	&	Adguard Content Blocker	&	com.adguard.android.contentblocker	&	Tools	&	4.06	&	Free	&	Browser	&	100K	\\
34	&	APUS Browser Turbo	&	com.apusapps.browser.turbo	&	Person.	&	4.23	&	Free	&	Browser	&	100K	\\
35	&	AdBlock for Samsung Internet	&	com.betafish.adblocksbrowser	&	Prod.	&	3.64	&	Free	&	Browser	&	100K	\\
36	&	Hermit • Lite Apps Browser	&	com.chimbori.hermitcrab	&	Tools	&	4.51	&	Free	&	Browser	&	100K	\\
37	&	CLIQZ Browser + Search Engine	&	com.cliqz.browser	&	Comm. 	&	4.04	&	Free	&	Browser	&	100K	\\
38	&	Anonymous Private Sec Browser	&	com.evda.connecttor	&	Comm. 	&	3.88	&	Free	&	Browser	&	100K	\\
39	&	Flyperlink	&	com.flyperinc.flyperlink	&	Prod.	&	4.15	&	Free	&	Browser	&	100K	\\
40	&	Jet Browser	&	com.jet.browser	&	Comm. 	&	4.08	&	Free	&	Browser	&	100K	\\
41	&	Browser	&	com.mmbox.browser	&	Comm. 	&	4.49	&	Free	&	Browser	&	100K	\\
42	&	X Browser	&	com.mmbox.xbrowser.gp	&	Tools	&	4.36	&	Free	&	Browser	&	100K	\\
43	&	X Browser  Super Fast \& mini	&	com.mmbox.xbrowser.pro	&	Tools	&	4.25	&	Free	&	Browser	&	100K	\\
44	&	WebGuard	&	com.mobisoft.webguard	&	Tools	&	4.54	&	Free	&	Browser	&	100K	\\
45	&	Adblock Fast	&	com.rocketshipapps.adblockfast	&	Prod.	&	3.87	&	Free	&	Browser	&	100K	\\
46	&	Swift Browser 	&	dotc.mobi.swift.browser	&	Comm. 	&	4.3	&	Free	&	Browser	&	100K	\\
47	&	3G - 4G Fast Internet Browser	&	jad.fast.internet.browser	&	Prod.	&	4.12	&	Free	&	Browser	&	100K	\\
48	&	AC Browser	&	net.ac.browser	&	Comm. 	&	4.53	&	Free	&	Browser	&	100K	\\
49	&	Rocket Browser	&	net.rocket.browser	&	Comm. 	&	4.44	&	Free	&	Browser	&	100K	\\
50	&	Fast Browser experience	&	sairam.simplebrowser	&	Comm. 	&	4.14	&	Free	&	Browser	&	100K	\\
51	&	Aon Browser	&	tr.abak.simsekTarayici	&	Comm. 	&	4	&	Free	&	Browser	&	100K	\\
52	&	LostNet NoRoot Firewall	&	com.lostnet.fw.free	&	Prod.	&	4.38	&	Free	&	VPN	&	50K	\\
53	&	Secure Wireless	&	me.disconnect.securefi	&	Tools	&	4.02	&	Free	&	VPN	&	50K	\\
54	&	Dee Browser	&	co.zew.deebrowser	&	Comm. 	&	4.14	&	Free	&	Browser	&	50K	\\
55	&	Adskip Browser	&	com.adsbrower	&	Tools	&	3.95	&	Free	&	Browser	&	50K	\\
56	&	Javelin Incognito Browser	&	com.jerky.browser2	&	Prod.	&	4.03	&	Free	&	Browser	&	50K	\\
57	&	\begin{CJK}{UTF8}{mj}팀버 애드필터 (Adfilter)\end{CJK}&	com.mobligation.timberadblock	&	Prod.	&	3.62	&	Free	&	Browser	&	50K	\\
58	&	\begin{CJK}{UTF8}{mj}팀버 브라우저 (Timber browser)\end{CJK}	&	com.mobligation.timberbrowser	&	Prod.	&	4.75	&	Free	&	Browser	&	50K	\\
59	&	YuBrowser - Fast, Filters Ads	&	com.mokee.yubrowser	&	Comm. 	&	4.1	&	Free	&	Browser	&	50K	\\
60	&	Slimperience Browser (AdBlock)	&	com.saschaha.browser	&	Comm. 	&	4.14	&	Free	&	Browser	&	50K	\\
61	&	Zirco Browser	&	org.zirco	&	Comm. 	&	3.78	&	Free	&	Browser	&	50K	\\
62	&	Dash Net Accelerated VPN	&	com.actmobile.dashnet	&	Travl.	&	3.95	&	Free	&	VPN	&	10K	\\
63	&	Adskip | for data control	&	com.play.adskip	&	Tools	&	2.23	&	Free	&	VPN	&	10K	\\
64	&	Lightning Web Browser +	&	acr.browser.lightning	&	Comm. 	&	4.32	&	1.50\$	&	Browser	&	10K	\\
65	&	AdBlocker Lite Browser	&	adblocker.lite.browser	&	Comm. 	&	3.66	&	Free	&	Browser	&	10K	\\
66	&	Crystal Adblock for Samsung	&	co.crystalapp.crystal	&	Prod.	&	4.05	&	Free	&	Browser	&	10K	\\
67	&	Secure Browser + Adblocker	&	com.browser.securebrowser	&	Prod.	&	3.97	&	Free	&	Browser	&	10K	\\
68	&	jioweb 4g browser	&	com.oorweb.uc.Activity	&	Comm. 	&	3.93	&	Free	&	Browser	&	10K	\\
69	&	Firebird Browser - Super Fast	&	com.prbstudios.firebird	&	Comm. 	&	3.98	&	Free	&	Browser	&	10K	\\
70	&	Clean Page - Adblocker Browser	&	com.tako.android.cleanpage	&	Comm. 	&	3.29	&	Free	&	Browser	&	10K	\\
71	&	No Ad Internet Browser	&	jad.internet.browser.block	&	Tools	&	2.59	&	Free	&	Browser	&	10K	\\
72	&	Unicorn Adblocker	&	kr.co.lylstudio.unicorn	&	Prod.	&	4.4	&	2.14\$	&	Browser	&	10K	\\
73	&	Trim Browser - Fast \& Secure	&	trim.altict.com	&	Comm. 	&	4.12	&	Free	&	Browser	&	10K	\\
74	&	PowerDownloads	&	zerolab.android.pdown	&	Video.	&	4.36	&	Free	&	Browser	&	10K	\\
75	&	sFly Network Booster	&	com.cdnren.sfly	&	Tools	&	4.1	&	Free	&	VPN	&	5K	\\
76	&	AdTrap Mobile	&	com.bluepointsecurity.adtrapaurora	&	Comm. 	&	3.07	&	Free	&	Browser	&	5K	\\
77	&	LostNet NoRoot Firewall Pro	&	com.lostnet.fw.pro	&	Prod.	&	4.15	&	0.99\$	&	VPN	&	1K	\\
78	&	MiniBrowser PRO	&	com.dmkho.mbm	&	Comm. 	&	4.28	&	1.50\$	&	Browser	&	1K	\\
79	&	Simple FLV with Adblock	&	com.titlisapp.simpleflv	&	Video.	&	4.52	&	Free	&	Browser	&	1K	\\
80	&	Deep Search Browser	&	com.wDeepSeacrhBrowser	&	Prod.	&	2.75	&	Free	&	Browser	&	1K	\\
81	&	Faster Browser Ever	&	com.wFastWebbrowserFree	&	Comm. 	&	4.09	&	Free	&	Browser	&	1K	\\
82	&	Supreme - Web Browser	&	es.soguezvm.supreme	&	Comm. 	&	3.94	&	Free	&	Browser	&	1K	\\
83	&	MSK Browser	&	gl.msk.bro	&	Comm. 	&	4.24	&	Free	&	Browser	&	1K	\\
84	&	Snap Browser	&	io.snapbrowser.lite	&	Comm. 	&	3.58	&	Free	&	Browser	&	1K	\\
85	&	Piggy browser	&	jp.myumyu.piggybrowser	&	Comm. 	&	3.62	&	Free	&	Browser	&	1K	\\
86	&	speed browser	&	siy.browser	&	Comm. 	&	4.1	&	Free	&	Browser	&	1K	\\
87	&	Krypton Browser	&	co.kr36.krypton.r	&	Comm. 	&	4.01	&	2.56\$	&	Browser	&	1K	\\
88	&	AndGuard for Root	&	soapbox.sym3try.andguard	&	Tools	&	3.85	&	1.95\$	&	VPN	&	500	\\
89	&	AndGuard Pro (w/ Iptables)	&	soapbox.sym3try.andguardpro	&	Tools	&	4.17	&	2.93\$	&	VPN	&	500	\\
90	&	Web Cleaner	&	com.chisstech.android2	&	Tools	&	4	&	Free	&	VPN	&	100	\\
91	&	Fast Speed Internet Browser	&	com.kubopirez.fast.speed.internet.browser	&	Tools	&	3.75	&	Free	&	Browser	&	100	\\
92	&	FastCat	&	com.wFastCat	&	Entert.	&	3.67	&	Free	&	Browser	&	100	\\
93	&	Magneto browser	&	com.wMagnetobrowser	&	Comm. 	&	3.94	&	Free	&	Browser	&	100	\\
94	&	Zeebrow	&	com.zohib.browser	&	Comm. 	&	3	&	Free	&	Browser	&	100	\\
95	&	Snap Browser Pro	&	io.snapbrowser.pro	&	Comm. 	&	4.1	&	1.58\$	&	Browser	&	100	\\
96	&	Browser Incognito - Supreme	&	es.soguezvm.incognito	&	Comm. 	&	5	&	3.12\$	&	Browser	&	50	\\
97	&	Supreme Pro - Web Browser	&	es.soguezvm.multi\_navegador\_pro	&	Comm. 	&	3.8	&	1.50\$	&	Browser	&	50	\\\hline
\bottomrule
\\
\caption{List of analyzed Android Adblocking apps with metadata as of December 20, 2016, on Google Play Store. }

\label{tab:alladblockers}
\end{longtable}
\twocolumn

\end{document}